\documentclass[10pt,journal,compsoc]{IEEEtran}
\usepackage{multirow}
\usepackage{subfigure}
\usepackage{caption}
\usepackage{verbatim}
\usepackage{etoolbox}
\usepackage{graphicx}
\usepackage{tcolorbox}
\usepackage{colortbl}
\usepackage{url} 
\usepackage{amsmath} 
\usepackage{hyperref}
\AtBeginEnvironment{table}{\small}
\usepackage{multirow}
\usepackage{balance}

\begin{document}

\title{Parameter-Efficient Fine-Tuning with Attributed Patch Semantic Graph for Automated Patch Correctness Assessment}

\author{Zhenyu~Yang,
        Jingwen~Wu,
        Zhen~Yang
        and Zhongxing~Yu
\IEEEcompsocitemizethanks{\IEEEcompsocthanksitem All the authors of this manuscript are with Shandong University, China.
E-mail: yangzycs@mail.sdu.edu.cn, elowen.jjw@gmail.com, zhenyang@sdu.edu.cn, and zhongxing.yu@sdu.edu.cn
}
\thanks{Manuscript received ...; revised ....}}


\IEEEtitleabstractindextext{%
\begin{abstract}
Automated program repair (APR) aims to automatically repair program errors without human intervention, and recent years have witnessed a growing interest on this research topic. While much progress has been made and techniques originating from different disciplines have been proposed, APR techniques generally suffer from the patch overfitting issue, i.e., the generated patches are not genuinely correct despite they pass the employed tests. To alleviate this issue, many research efforts have been devoted for automated patch correctness assessment (APCA). In particular, with the emergence of large language model (LLM) technology, researchers have employed LLM to assess the patch correctness and have obtained the state-of-the-art performance. The literature on APCA has demonstrated the importance of capturing patch semantic and explicitly considering certain code attributes in predicting patch correctness. However, existing LLM-based methods typically treat code as token sequences and ignore the inherent formal structure for code, making it difficult to capture the deep patch semantics. Moreover, these LLM-based methods also do not explicitly account for enough code attributes. To overcome these drawbacks, we in this paper design a novel patch graph representation named attributed patch semantic graph (APSG), which adequately captures the patch semantic and explicitly reflects important patch attributes. To effectively use graph information in APSG, we accordingly propose a new parameter-efficient fine-tuning (PEFT) method of LLMs named Graph-LoRA. Our method focuses on the typical real-world scenarios where ground-truth patches are inaccessible, and does not rely on ground-truth patches to work. 
Extensive evaluations have been conducted to evaluate our method, and the results show that compared to the state-of-the-art methods, our method improves the accuracy and F1 score by 3.1\% to 7.5\% and 3.0\% to 7.1\% respectively.

\end{abstract}

}

\maketitle
\IEEEdisplaynontitleabstractindextext
\IEEEpeerreviewmaketitle
\ifCLASSOPTIONcompsoc
\IEEEraisesectionheading{\section{Introduction}\label{sec:introduction}}
\else
\section{Introduction}
\label{sec:introduction}
\fi

\noindent Software is unfortunately plagued with bugs, which can have serious consequences such as data loss, security flaw, and system hang. Resolving the software bugs is a notoriously difficult, expensive, and error-prone process \cite{debugging,multiple-fault}, and this issue is getting more and more serious as the scale and complexity of software continue to expand. To alleviate the burden on developers, the area of automated program repair (APR) arises and has received widespread attention from both academia and industry in the past two decades \cite{urli2018design,9393494}. The research agenda of APR is to automatically fix bugs in programs with less human intervention, and techniques originating from different disciplines have been proposed, remarkably including heuristic repair \cite{le2011genprog, yuan2018arja, xin2019better, kim2013automatic}, constraint-based repair \cite{xuan2016nopol,nguyen2013semfix,mechtaev2015directfix}, and learning-based repair \cite{chen2019sequencer, ye2022neural, Recoder, yutse}. 

Roughly speaking, APR techniques consist of three phases: fault localization \cite{7390282,YUgui,6113915}, patch generation \cite{6606626}, and patch validation \cite{9402121}. For the patch validation phase, the proposed techniques within the APR community typically evaluate the correctness of the generated patches using manually written test cases and a patch is deemed as correct in case it cam make the program pass all test cases. However, test cases in general can not fully specify the program behaviors and existing studies \cite{overfittingsurvey, qi2015analysis, long2016analysis, yuemse,smith2015cure,10.1145/3663529.3663776,10.1145/3533767.3534368} demonstrate the existence of a significant portion of patches which pass the existing test suite but are actually incorrect. This phenomenon is called the patch overfitting problem, meaning that the generated patches simply overfit the existing test suite but do not achieve the expected program behavior in general. 

To alleviate this serious issue which overshadows the APR area, researchers have proposed many techniques for automated patch correctness assessment (APCA) \cite{tian2020evaluating,ye2021automated,xiong2018identifying,smith2015cure,le2023invalidator, tian2022predicting,ye2021comprehensive}. Currently, APCA techniques can be broadly divided into two categories  \cite{overfittingsurvey}: dynamic methods and static methods. Dynamic methods determine the correctness of patches by generating additional test cases and/or collecting features of test execution. For example, \emph{Opad} \cite{yang2017better} makes use of fuzz testing to generate new test cases and employs the corresponding oracles to enhance the patch correctness verification. Dynamic methods can achieve high accuracy but are very time-consuming due to the generation and/or running of tests. The static approach does not rely on running tests and instead assesses the patch correctness by the characteristics of the patch, such as its syntax and static semantic. For instance, on top of the assumption that the correct patch should be more similar to the defective code, \emph{S3} \cite{le2017s3} measures the syntactic and semantic similarities between the patch and the error code to assess the patch correctness. Compared with dynamic approaches, static approaches can assess patch correctness quickly but suffer from the prediction accuracy issue. An ideal APCA approach should simultaneously maintain low time cost and high accuracy. 

Given that dynamic approaches necessarily take time to generate and/or run tests, much research attention has been devoted to static approaches in recent years and improving the accuracy of static approaches is viewed as a breakthrough \cite{lin2022context,zhou2024leveraging}. With the development of machine learning techniques, numerous learning-based APCA methods in the static category have been proposed in recent years. For instance, Ye et al. \cite{ye2021automated} manually design and extract 202 code features from the abstract syntax trees of defective code and patch code. Then, these code features and labels are given as inputs to three machine-learning models for constructing a probabilistic model. 
More recently, enlightened by the remarkable success of large language models (LLMs) for promoting code intelligence \cite{10.1145/3695988,10713474}, some researchers have employed LLMs to statically assess the correctness of patches. 
In particular, Zhou et al. \cite{zhou2024leveraging} propose LLM4PatchCorrect, which predicts the patch correctness by feeding an LLM with information of labeled patches, such as error descriptions and failed tests. 
While these LLM-based methods have achieved state-of-the-art results in statically predicting patch correctness, drawbacks are associated with them. Previous works have demonstrated the importance of capturing patch semantic (including semantics of both the changed code and related unchanged code) in statically predicting patch correctness \cite{wen2018context,lin2022context}, and an abundance of works also have shown that explicitly considering certain 
attributes associated with the changed and related unchanged code are extremely beneficial \cite{ye2021automated,le2017s3,xia2023automated}. However, existing LLM-based methods typically treat code as token sequences and ignore the inherent formal structure for code, making it difficult to capture the deep patch semantics. Moreover, these LLM-based methods also do not explicitly account for enough attributes associated with the changed and unchanged code. Overall, these two drawbacks lead to the degraded performance in statically predicting patch correctness for LLM-based methods.

To overcome the drawbacks, we in this paper propose a novel patch graph representation named \emph{Attributed Patch Semantic Graph} (APSG). APSG is a directed graph which not only adequately captures the patch semantic through data and control flow between program elements, but also captures important attributes associated with the changed and related unchanged code by labeling different types of APSG nodes with different types of explicit attributes. Upon generating APSG, we further incorporate information of APSG into LLMs for statically predicting patch correctness. Inspired by the work of Yao et al. \cite{yao2020multimodal}, we find that the attention mechanism can effectively merge graph information in APSG with sequence information in LLM. Besides, LLMs need to be fine-tuned in order to adapt to the APCA task. Taking these aspects into account, on top of LoRA \cite{hu2021lora}---one of the most advanced LLM parameter-efficient fine-tuning (PEFT) methods, we propose a new PEFT method called Graph-LoRA to retain graph information and fully train LLMs. Graph-LoRA can effectively fine-tune the parameters of LLMs and incorporate APSG information into LLMs through the attention mechanism. Our method focuses on the typical real-world scenarios where ground-truth patches are inaccessible, and does not rely on ground-truth patches to work.

To verify the effectiveness of our method, we conduct experiments on five APCA datasets, including the Wang dataset \cite{wang2020automated}, the Merge dataset \cite{yang2023large}, the Balance dataset \cite{yang2023large}, the Lin dataset \cite{lin2022context}, and the Multi-Benchmarks dataset. The first four datasets are based on the Defects4J benchmark \cite{just2014defects4j}, and have been widely used in the APCA field. The Multi-Benchmarks dataset is a large dataset we built that involves three bug benchmarks: Defects4J, Bugs.jar \cite{saha2018bugs}, and Bears \cite{madeiral2019bears}. The experimental results show that our method outperforms all static methods, including traditional methods and learning-based methods, in terms of accuracy, precision, recall, and F1 score. Compared with the best static method LLM4PatchCorrect \cite{zhou2024leveraging}, our method improves the accuracy and F1 score by 3.1\% to 7.5\% and 3.0\% to 7.1\% respectively. 
In terms of precision, our method is closest to the dynamic method Opad \cite{yang2017better}. The results of ablation studies show that both APSG and Graph-LoRA play a significant role in fine-tuning LLM on the APCA task. The results of cross-project prediction also show that our method achieves state-of-the-art performance when evaluating unseen patches. 


In summary, our primary contributions are as follows:
\begin{itemize}
    \item We propose a novel patch graph representation named Attributed Patch Semantic Graph (APSG), which not only adequately captures the patch semantic but also explicitly reflects important attributes associated with the patch. 
    \item We propose a new parameter-efficient fine-tuning method of LLMs named Graph-LoRA, which can effectively incorporate additional graph information while fine-tuning LLMs. 
    \item We conduct large-scale experimental evaluations and the results clearly show that our approach outperforms the state-of-the-art in automated patch correctness assessment. 
\end{itemize}

Our replication package (including code, dataset, etc.) is available at \url{https://github.com/SEdeepL/GraphLoRA}.

\section{Related Work} 
This section reviews existing literature closely related to our work in this paper, including literature on automated patch correctness assessment and literature on LLM and PEFT.

\subsection{Automated Patch Correctness Assessment.} 
In the area of automated program repair (APR), there are typically two approaches for assessing the correctness of generated patches. The first approach is manual annotation where the correctness of generated patches is determined manually, and the second approach is automated assessment where manual efforts are not required. 
The study by Le et al. \cite{le2019reliability} shows that manual annotation is more effective compared to automated assessment, but involves significant costs. Consequently, recent research efforts have been devoted primarily to
automated patch correctness assessment (APCA), aiming to improve the effectiveness of APCA methods.
Depending on whether test case executions are needed, these APCA methods can be broadly divided into dynamic methods and static methods \cite{overfittingsurvey}. 

Dynamic methods determine the correctness of patches by making use of test case generation techniques \cite{47977,6319229} (particularly test amplification techniques that generate additional test cases \cite{danglot2017emerging}) and/or collecting features of test execution \cite{shamshiri2015automatically}. Yu et al. \cite{yuemse} give the definition of two different kinds of overfitting issues, that is, \emph{incomplete fixing} and \emph{regression introduction}, and the proposed dynamic methods typically have different strengths for them.
Xin et al. \cite{xin2017identifying} propose DiffTGen, which uses the test generation tool Evosuite \cite{10.1145/2025113.2025179} to generate additional test cases for enhancing the patch correctness check. PATCH-SIM, proposed by Xiong et al. \cite{xiong2018identifying} , uses a test generation tool to generate new test cases and assesses the correctness of the patch based on the similarity of the test case execution. The underlying assumption is that a correct patch should make the original program and the patched program behave similarly on the test cases that originally passed, but behave differently on the test cases that originally failed. Yang et al. \cite{yang2017better} propose using fuzz testing to generate additional test cases and setting corresponding test assertions to validate the correctness of patches.

In contrast, the static methods do not need to run test cases and the correctness of the patch is assessed by the characteristics of the patch. Le et al. \cite{le2017s3} assume that the correct patch should be more similar to the defective code and propose S3 based on this assumption. On top of six features, S3 measures the syntactic and semantic similarities between the patch and the defective code to assess the patch correctness. 
Wen et al. \cite{wen2018context} focus more on the contextual information of the patch and design three context-aware functions to assess the patch correctness. Xia et al. \cite{xia2023automated} first introduce entropy into the APCA task. They assume that the correct patches are more natural than the overfitting patches, and the entropy of patches can be used to measure patch correctness. With the development of machine learning techniques, many researchers have developed learning-based APCA methods. Ye et al. \cite{ye2021automated} manually design and extract 202 code features from the abstract syntax trees of the defective code and patch code. These code features and labels are then given as inputs to three machine-learning models for constructing a probabilistic model. Csuvik et al. \cite{csuvik2020utilizing} attempt to use BERT to generate embedding vectors to determine the similarity between the defective code and the patch code, thereby filtering overfitting patches. Zhang et al. \cite{zhang2024appt} use the BERT \cite{kenton2019bert} model as the encoder stack and use LSTM \cite{hochreiter1997long} to determine patch correctness. Tian et al. \cite{tian2022change} use a neural network architecture to learn the semantic correlation between the bug reports and code patches to measure patch correctness. Tian et al. \cite{tian2022predicting} additionally propose BATS, which predicts patch correctness based on the similarity of failed test cases. Lin et al. \cite{lin2022context} use contextual and structural information to modify patch embeddings for improving the accuracy of patch correctness assessment. 

Most recently, enlightened by the remarkable success of large language models (LLMs) for promoting code intelligence, some researchers
have employed LLMs to statically assess the correctness of patches. Notably, Zhou et al. \cite{zhou2024leveraging} predict the patch correctness by feeding an LLM with information of labeled patches, such as error descriptions and failed tests.

Currently, dynamic methods have limited practical applications due to the disadvantage of requiring a lot of time to generate and/or execute test cases. Static methods suffer from the accuracy issue. LLMs have achieved remarkable performance in code intelligence and may be a breakthrough in improving the performance of static methods. However, existing LLM-based methods ignore the inherent formal structure for code and do not explicitly account for enough attributes associated with the changed and unchanged code, leading to degraded prediction performance.

\subsection{Large Language Model and Parameter-Efficient Fine-Tuning.} With the continuous development of deep learning technology and computing power, researchers have proposed various LLMs. In 2022, Google released an LLM called Chat-GPT \cite{chatgpt}, which demonstrates outstanding performance in the question-answering field. Touvron et al. \cite{touvron2023llama} train an LLM called LLama using only public data, and release the model parameters to the research community. LLama has become the most popular open-source LLM. Roziere et al. \cite{roziere2023code} propose an open-source LLM for code, which has significantly improved performance for numerous tasks, including content filling, context information extraction, and instruction tracking. Li et al. \cite{li2023starcoder} propose another open-source LLM for code named StarCoder, which expands the model input length to 8K and demonstrates excellent performance in the Python language.

\begin{figure*}
\centering
\includegraphics[width=0.85\textwidth]{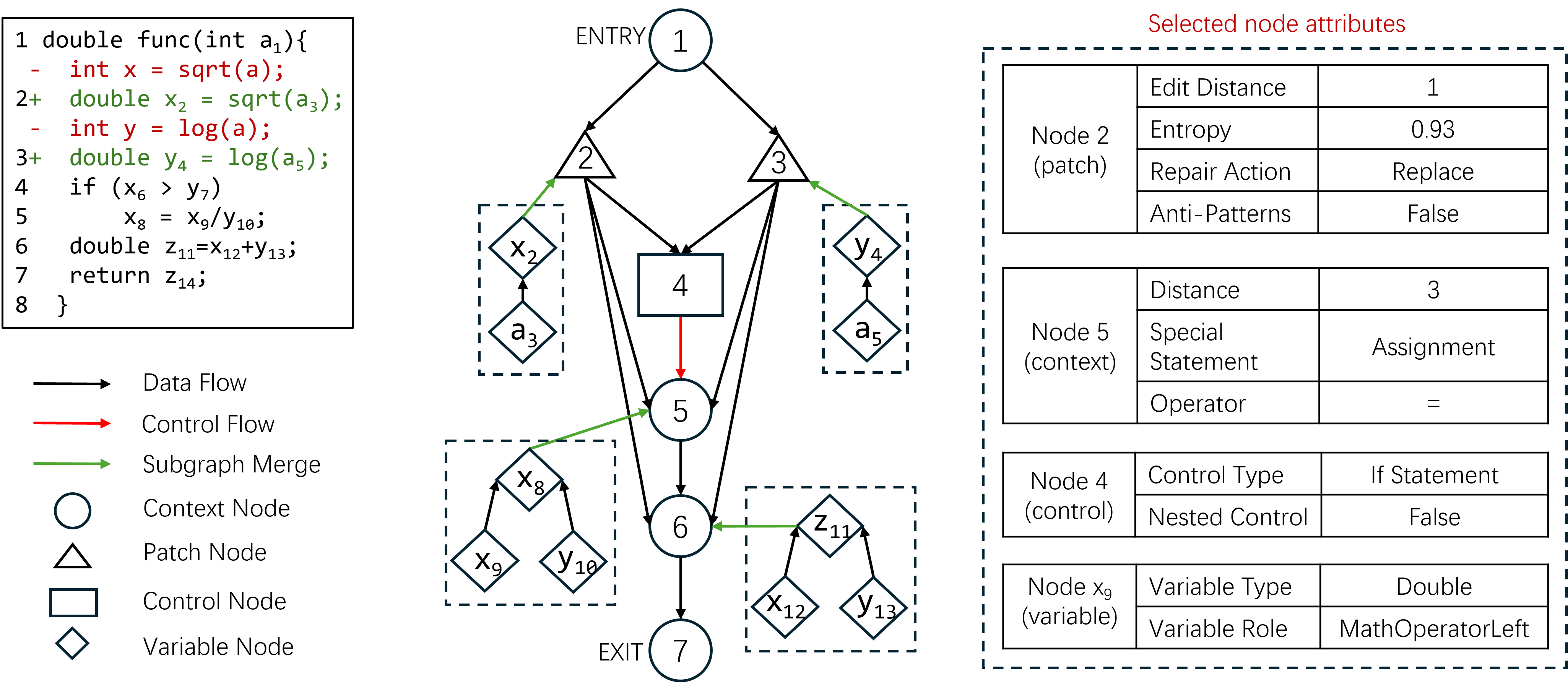}
\caption{\label{fig:frog1}An example of the attributed patch semantic graph.}
\end{figure*}

As LLMs become more common, it is particularly important to optimize computational efficiency and resource usage. The purpose of parameter-efficient fine-tuning (PEFT) is to reduce resource consumption during fine-tuning by training only a part of the model's parameters. Houlsby et al. \cite{houlsby2019parameter} add an adapter module to each layer of the pre-trained model, froze the main parameters of the model during fine-tuning and only fine-tune the newly added adapter structure.
Inspired by the concept of prompt, Li et al. \cite{li2021prefix} propose prefix tuning, another fine-tuning method based on adding parameters. The method constructs a continuous and task-related prefix and only modifies the prefix of a specific task during model training. Guo et al. \cite{guo2020parameter} propose a parameter-modifying fine-tuning method called Diffpruning, which describes fine-tuning as learning a diff vector and adding it to the pre-trained fixed model parameters. Hu et al. \cite{hu2021lora} assume that the model parameters can be updated by modifying the intrinsic rank of the parameters, and further propose an intrinsic rank adapter LoRA for fine-tuning LLMs. Based on LoRA, Chen et al. \cite{chen2023longlora} propose LongLoRA, which splits the long context and processes each group of context separately through the shifted sparse attention mechanism.

For the excellent performance of LLMs, we aim to use LLMs to alleviate the accuracy problem of static patch correctness evaluation methods.

\section{Attributed Patch Semantic Graph}\label{psg}
This section introduces the Attributed Patch Semantic Graph (APSG), a novel patch graph representation which aims to facilitate LLM-based methods for statically predicting patch correctness. 

With the development of deep learning techniques, a significant portion of research efforts have been devoted to learning-based code intelligence and impressive results have indeed been obtained. One key to the success of these learning-based methods lies in appropriate code representation \cite{surveycode}. Currently, there are three main types of code representation methods within the literature: token-based method \cite{Miltiadis2,Jinghui}, syntax-based method \cite{code2vec,Michael}, and semantic-based method \cite{Jakobovits,DependencyGraph,Kwonsoo,graphcodebert}. Token-based methods represent the code as a series of tokens, and this simple representation facilitates learning but limited semantics can be captured due to the ignorance of the inherent code structure. Syntax-based methods represent code in the form of trees, which contain rich semantic information but usually have a deep hierarchical structure. As a result, in practice, notable refinement efforts of the raw tree representation are typically required to make the learning a success. Semantic-based methods represent code in the form of graphs, which can effectively facilitate the capture of code semantics for learning models. Among the variety of proposed graph representations, notable ones include data flow graph \cite{Kwonsoo}, control flow graph \cite{david2020neural}, program dependence graph \cite{DependencyGraph}, and contextual flow graph \cite{Jakobovits}.

In the APCA field, previous works have demonstrated the importance of capturing patch semantic (including semantics of both the changed code and related unchanged code) in statically predicting patch correctness \cite{wen2018context,lin2022context}, and an abundance of works also have shown that explicitly considering certain attributes associated with the changed and related unchanged code are extremely beneficial \cite{ye2021automated,le2017s3,xia2023automated}. While existing graph-based representations can effectively facilitate the code semantic learning, they typically do not simultaneously contain changed and unchanged code. In addition, existing graph-based representations do not involve any explicit code attributes. In light of these shortcomings, we in this paper design a novel directed patch graph representation named Attributed Patch Semantic Graph (APSG). For ease of reading, {Fig.~\ref{fig:frog1} gives an example of an Attributed Patch Semantic Graph for a simple patch. 

\vspace{0.6mm}
\textbf{Definition (Attributed Patch Semantic Graph)} The Attributed Patch Semantic Graph for a patch is a triple tuple $<V, E, X>$ where $V$ is a set of nodes, $E$ is a set of directed edges between nodes in $V$, and $X$ is a mapping from nodes in $V$ to their attributes.
\vspace{0.6mm}

We next give a detailed explanation of the graph. First, the node set $V$ can be further divided into four categories: patch node set $V_p$, control node set $V_c$, context node set $V_{ct}$, and variable node set $V_v$. As a single patch is typically viewed as a collection of statement-level code changes, the patch node set $V_p$ corresponds to the set of changed code statements for the patch. Accordingly, the control node set $V_c$ and context node set $V_{ct}$ correspond to the set of surrounding control statements and the set of surrounding non-control statements respectively. Our current analysis unit is a method, so the set of statements involved with $V_p$, $V_c$, and $V_{ct}$ are within a method body. In particular, if the method declaration involves parameters, there is a special entry statement node which essentially corresponds to a variable declaration statement. The node labeled with 1 in {Fig.~\ref{fig:frog1} is an example of this special node. The variable node set $V_v$ corresponds to the set of involved variables in assignment statements (including statements which simultaneously contain declaration and assignment, like statement 6 in Fig.~\ref{fig:frog1}), and is introduced for capturing more code semantics (explained more in the next point). 
\begin{table*}[]
\centering
\caption{\label{tab:t1}The list of considered node attributes in APSG.}
\resizebox{0.85\textwidth}{!}{
\begin{tabular}{lll}
\hline
Node Type                      & Attribute Type         & Attribute  Content                                                                                                    \\ \hline
\multirow{4}{*}{Patch node}    & Edit Distance          & Manhattan distance between the defective code and the patch                                                           \\ \cline{2-3} 
                               & Entropy Score          & Code line entropy score                                                                                               \\ \cline{2-3} 
                               & Repair Action          & Addition, Deletion, and Replacement                                                                                   \\ \cline{2-3} 
                               & Anti-pattern           & Whether the patch conforms to anti-patterns                                                                           \\ \hline
\multirow{3}{*}{Context node}  & Distance to Patch      & The distance from the node to the patch in APSG                                                                       \\ \cline{2-3} 
                               & Special Statement Type & Assignment, Try-catch, Invocation, and Return                                                                         \\ \cline{2-3} 
                               & Operator Type          & \begin{tabular}[c]{@{}l@{}}Binary-Operator, Unary-Operator, \\ Relational-Operator, and Bitwise-Operator\end{tabular} \\ \hline
\multirow{2}{*}{Control node}  & Control Type           & If statement, Switch statement, While statement, and For statement                                                    \\ \cline{2-3} 
                               & Nested Control         & Whether the control statement is a nested control                                                                     \\ \hline
\multirow{2}{*}{Variable node} & Variable Type          & The type of the variable in the code                                                                                  \\ \cline{2-3} 
                               & Variable Role          & The role of the variable in computation                                                                               \\ \hline
\end{tabular}
}
\end{table*}
Second, a directed edge in $E$ can be of 3 kinds: data flow edge, control flow edge, and sub-graph merge edge. The data flow and control flow edges established between statement nodes (i.e., nodes from the sets $V_p$, $V_c$, and $V_{ct}$) are similar to that of the typical program dependence graph. 
For control flow, there exists a control flow edge from node $a$ to node $b$ if $a$ represents the conditional statement whose predicate outcome directly controls whether $b$ is executed (\emph{the edge from node labeled with 4 to node labeled with 5 in Fig.~\ref{fig:frog1} is an example}). For data flow, there exists a data flow edge from node $a$ to node $b$ in case a certain variable $v$ defined at $a$ is used at $b$ and there is a path of the form $a$ · $P$ · $b$, where $P$ is a path along which $v$ is not redefined (\emph{the edge from node labeled with 2 to node labeled with 4 in Fig.~\ref{fig:frog1} is an example, the involved variable is $x$}). Previous studies \cite{graphcodebert,StructCoder} have demonstrated the significance of considering data flow inside statements for accurately capturing code semantics, we thus also consider this aspect in APSG. In particular, there exists a data flow edge from node $a$ (in set $V_v$) to node $b$ (in set $V_v$) in case $a$ and $b$ correspond to variables on the right and left sides of an assignment statement respectively, and there exists a sub-graph merge edge from node $a$ (in set $V_v$) to node $b$ if $a$ corresponds to a variable on the left side of an assignment statement and $b$ corresponds to the assignment statement (\emph{the sub-graph for node labeled with 5 in Fig.~\ref{fig:frog1} is an example}). 

Finally, the mapping $X$ maps each node in the set $V$ to certain attributes. As nodes in $V$ are diverse, we consider different attributes for different node categories. Most of the node attributes are adapted from the relevant literature, and Table \ref{tab:t1} lists all the considered attributes.

\begin{itemize}
\item The \emph{patch node attribute} is used to describe the characteristics of the patch at the line level, and we have considered four attributes for patch nodes. The first attribute is the edit distance, which calculates the number of times required for editing the defective code into the patch code based on the Manhattan distance. The second attribute is the patch entropy value, which describes the naturalness of a patch by calculating the maximum entropy of the patch and the calculation procedure follows that proposed by Xia et al. \cite{xia2023automated}. The third attribute is about the repair action, including addition, deletion, and replacement. The fourth property is the anti-pattern, which assesses whether the patch involves the forbidden transformations of the overfitting patches defined by Tan et al. \cite{tan2016anti}. 

\item The \emph{context node attribute} is used to describe the characteristics of the context related with the patch correctness, and we have considered three attributes for context nodes. The first attribute is the distance to the patch, which describes the importance of context nodes in APSG by calculating the distance from the context node to the patch. The second attribute is about special statement type, including assignment statement, try-catch statement, invocation statement, and return statement. The third attribute is about the operator type (if involved in the corresponding statement), including binary operator, unary operator, relational operator, and bitwise operator. 

\item The \emph{control node attribute} is used to describe the characteristics of the control statement, and we have considered two
attributes for control nodes. The first attribute is about special control statement type, including if statement, switch statement, while statement, and for statement. The second attribute is about whether the control statement belongs to the body of another control statement, i.e., whether it is a nested control. 

\item The \emph{variable node attribute} represents the characteristics of the variable, and we have considered two attributes for variable nodes. The first attribute is variable type, which establishes the specified type for the variable, such as int, float, double, and bool. The second attribute is variable role, which describes the role of variables in computation and
the calculation procedure follows that proposed by Du et al. \cite{yufse23}. As an example, in the code snippet ``\texttt{a + b}'', the roles of variables \texttt{a} and \texttt{b} are \texttt{MathOperatorLeft} and \texttt{MathOperatorRight}, respectively.

\end{itemize}



In summary, APSG is a directed graph which not only adequately captures the patch semantic through data and control flow between program elements, but also captures important attributes associated with the changed and related unchanged code by labeling different categories of APSG nodes with different types of explicit attributes. These merits make APSG a strong candidate graph representation for LLM-based patch correctness prediction methods. 

Based on the code analysis tool Spoon \cite{spoon}, we fully implement an analyzer to get APSG for a Java method and the analyzer supports modern Java versions up to Java 16. 

\section{Graph-LoRA for LLMs}
In this section, we describe our proposed parameter-efficient fine-tuning method named Graph-LoRA, which effectively incorporates APSG information into LLMs during fine-tuning to improve the performance of LLMs on the APCA task.
\begin{figure*}
\centering
\includegraphics[width=0.85\textwidth]{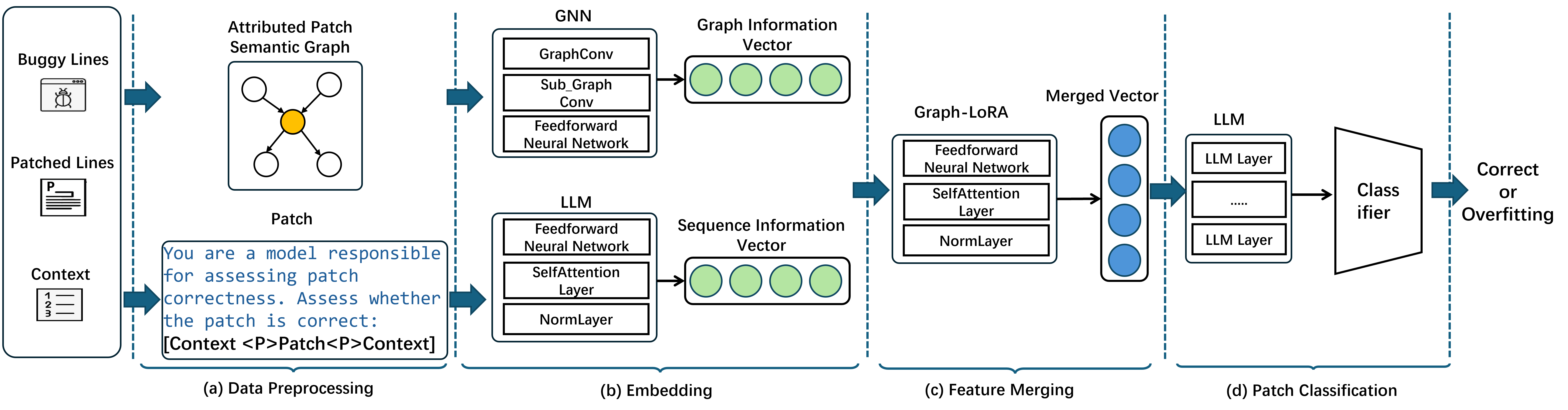}
\caption{\label{fig:frog2}An overview of fine-tuning LLM with Graph-LoRA in our method.}
\end{figure*}
\subsection{Overview}
\textbf{Motivation:}   
Previous works have demonstrated the importance of capturing patch semantic and explicitly considering certain code attributes in statically predicting patch correctness \cite{wen2018context,lin2022context,ye2021automated,le2017s3,xia2023automated}. While the proposed patch graph representation APSG adequately captures such information, we further need to effectively incorporate information of APSG into LLMs for statically predicting patch correctness. Inspired by the work of Yao et al. \cite{yao2020multimodal}, we find that the attention mechanism can effectively merge the graph information in APSG with the sequence information in LLMs. Besides, LLMs need to be fine-tuned in order to adapt to the APCA task. Taking these aspects into account, on top of LoRA \cite{hu2021lora}---one of the most advanced LLM parameter-efficient fine-tuning (PEFT) methods, we propose a new PEFT method called Graph-LoRA to retain graph information and fully train LLMs. Graph-LoRA can effectively fine-tune the parameters of LLMs and incorporate APSG information into LLMs through the attention mechanism.


\vspace{1.0mm}
\noindent\textbf{Framework:}
Fig.~\ref{fig:frog2} shows the process of fine-tuning LLMs with Graph-LoRA in our method. Given the buggy line(s), patched line(s) and context of the buggy code, the specific process of our method is as follows: (a) First, we preprocess the patch data to obtain the APSG and sequence representation of the patch; (b) Then, we obtain the graph features and sequence features of the patch through GNN and LLM respectively; (c) Next, we use the attention mechanism of Graph-LoRA to merge the graph features with sequence features; (d) Finally, LLM processes the merged features and assesses the correctness of the patch.

\subsection{Code Embedding for Sequence Features}
LLMs can convert the patch into token embeddings that will be used for prediction in subsequent modules. In this work, we use LLMs built by the stacked decoder of transformers \cite{vaswani2017attention}, which is the most popular LLM in the field of software engineering.

Given an input sequence $X$ of a code piece containing the patch, let $X_i$ be the i-th token of the code piece. To make the LLMs clearly distinguish the patch content, we use pre-appended token $<P>$ to wrap the beginning and end of the patch. In addition, we add the text "You are a model responsible for assessing patch correctness. Assess whether the patch is correct" in the front of the code piece, which serves as a prompt to LLM. Finally, the code piece with patch is represented as:
\begin{equation}
X = \{Prompt: \mathbf{x}_1, \ldots <P>,\mathbf{x}_m, \ldots, <P>, \ldots, \mathbf{x}_n\}
\end{equation}

Code tokens are then converted into fixed-dimensional vector representations via LLM, and the code vector is represented as:
\begin{equation}
E=Embedding(X) = \{\mathbf{e}_1, \mathbf{e}_2, \ldots, \mathbf{e}_n\}
\end{equation}
where $E$ represents the feature vector of this code piece and $\mathbf{e}_i$ is the feature vector of the i-th code token.
\subsection{GNN for Graph Features}
To enrich the feature vector of the patch, we need to additionally get the feature of the APSG. According to the procedure in Section~\ref{psg}, we can build the APSG of the patch and extract the node matrix $N$, adjacency matrix $M$, and attribute matrix $A$. The node matrix includes the line node matrix $N_l$ and the variable node matrix $N_v$, and the attribute matrix includes the attributes of each node. The adjacency matrix includes the line node adjacency matrix $M_l$, the variable node adjacency matrix $M_v$, and the sub-graph merge edge matrix $M_{l\_v}$. The line node matrix, line node adjacency matrix, and line node attribute matrix form the overall graph. The variable node matrix, variable node adjacency matrix, and variable node attribute matrix form the subgraph. To effectively obtain the graph information of APSG, given the powerful ability of the Graph Neural Network (GNN) \cite{scarselli2008graph}, we make use of GNN to process graph data and extract features of APSG. The process of extracting feature of APSG is as follows: (a) First, we process node features and node attribute features; (b) Then, based on the node and its attribute features, we extract subgraph features; (c) Finally, we pass the subgraph features to the overall graph and extract the overall graph features. We next give details of the three steps. 
\begin{figure*}
\centering
\includegraphics[width=0.75\textwidth]{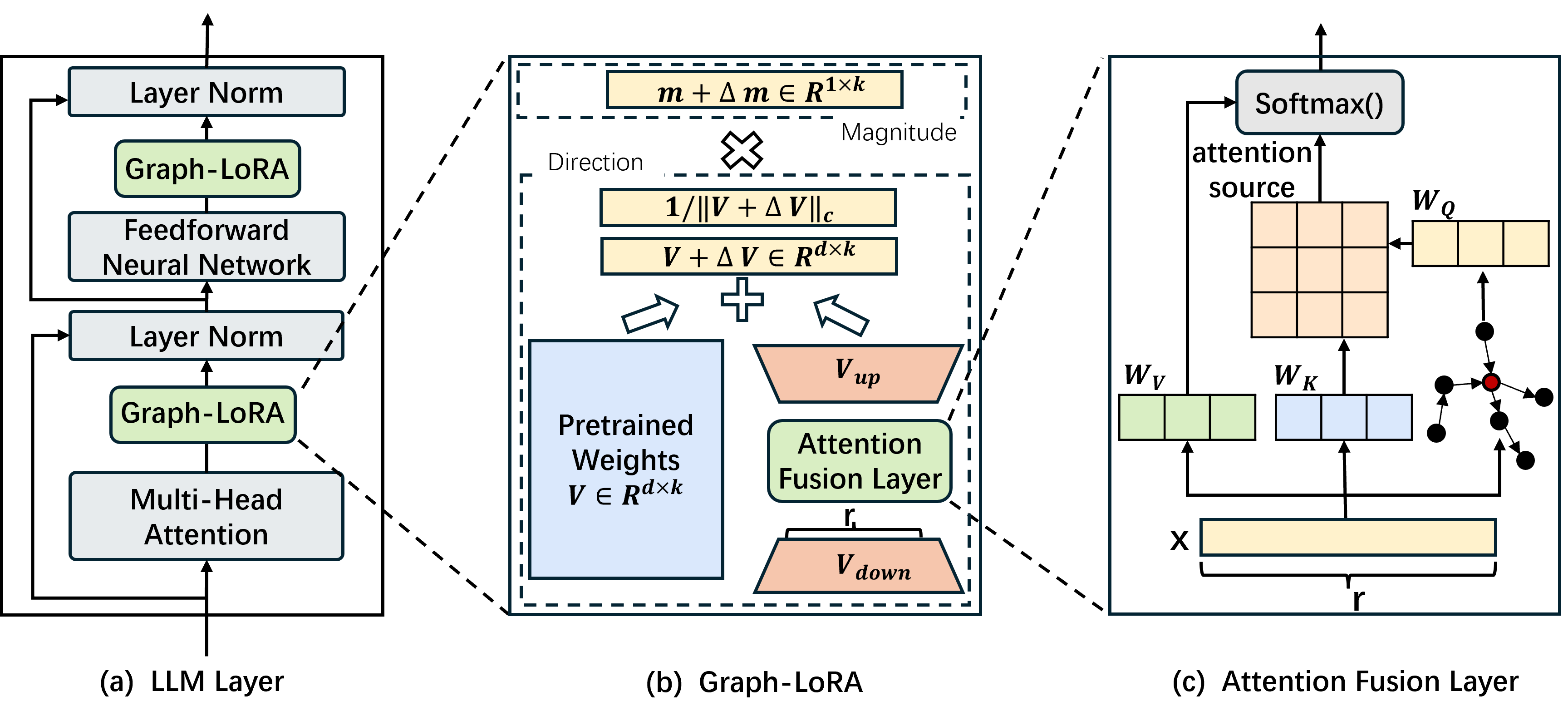}
\caption{\label{fig:frog3}An overview of Graph-LoRA. The left part is a layer of LLM, the middle part is the Graph-LoRA, and the right part is the attention fusion layer.}
\end{figure*}

First, we process nodes and attributes in APSG. To incorporate node attributes into graph information, we merge node attribute embedding and node embedding. The specific operations are as follows: 
\begin{equation}
    {H}_{n} = concat(N,A)
\end{equation}
\begin{equation}
    {F}_{n} = {Linear}_{1}({H}_{n})
\end{equation}
where ${Linear}_{1}$ is a feed-forward network layer used to change the node feature dimension. Following this approach, we get new line node features ${F}_{l}$ and new variable node features ${F}_{v}$.

Then, we need to obtain the features of the subgraph composed of variable nodes. To achieve this, we use graph convolution to aggregate node features within a subgraph. By passing node information, graph convolution can effectively obtain subgraph features. The specific operations are as follows: 
\begin{equation}
H_v = Sub\_GraphConv({F}_{v},M_v)
\end{equation}
\begin{equation}
Sub\_GraphConv({F}_{v},M_v) = \sigma\left( D_v^{-1/2} M_v D_v^{-1/2}{F}_{v} W_v \right)
\end{equation}
\begin{equation}
{D_v}=\sum_{i=1}^{n}{M}_{v}
\end{equation}
where $H_v$ is the feature of the subgraph composed of variable nodes, $D_v$ is the degree matrix of the variable node, $W_v$ is the weight matrix, and $\sigma$ is the nonlinear activation function. 

Finally, after obtaining the subgraph features, we aggregate the subgraph features into the line nodes to get features of the overall graph. According to the sub-graph merge edge matrix $M_{l\_v}$, we fuse subgraph features with corresponding line node features. We do not change line node features without subgraphs. Furthermore, we use graph convolution to get features of the overall graph. The specific operations are as follows: 
\begin{equation}
    {H}_{l} = {Linear}_{2}(Concat({F}_{l},H_v \cdot M_{l\_v}))
\end{equation}
\begin{equation}
    {H}_{out} = GraphConv(M_l,H_l)
\end{equation}
\begin{equation}
GraphConv(M_l,H_l) = \sigma\left( D_l^{-1/2} M_l D_l^{-1/2}H_l W_l \right)
\end{equation}
\begin{equation}
{D_l}=\sum_{i=1}^{m}{M}_{l}
\end{equation}
where $H_l$ is the line node feature matrix, ${H}_{out}$ is the feature of the APSG, ${Linear}_{2}$ is a feed-forward network layer that changes the line feature dimension, $D_l$ is the degree matrix of the line node, and $W_l$ is the weight matrix.
\subsection{Graph-LoRA}
After obtaining the APSG features, we need to make use of them to help LLMs determine the patch correctness more accurately. To achieve this, we propose Graph-LoRA, a novel parameter-efficient fine-tuning (PEFT) method that can incorporate graph information into LLMs. Fig.~\ref{fig:frog3} shows an overview of Graph-LoRA, and below we will introduce the specific process of Graph-LoRA. 



First, Graph-LoRA decomposes the pre-trained weights $W_0 \in R^{d \times k}$ of LLMs into a vector $m \in R^{1 \times k}$ representing ``magnitude'' and a matrix $V \in R^{d \times k}$ representing ``direction'' on top of DoRA \cite{liu2024dora}, an advanced variant of LoRA. Compared to the original LoRA method that fine-tunes all content in one step, this method allows LLMs to clearly know the magnitude and direction of the weights that need to be fine-tuned, resulting in better performance. For the magnitude vector, we fine-tune all its parameters using the weight update vector $\Delta m$. For the direction matrix, we decompose its update matrix $\Delta V$ into low-rank matrices $V_{down} \in R^{d \times r}$ and $V_{up} \in R^{r \times k}$ using low-rank decomposition. The specific operation is as follows:
\begin{equation}
m = \left\lVert W_0 \right\rVert_{c}  \quad\quad
V = W_0
\end{equation}
\begin{equation}
\Delta V = V_{down} V_{up}
\end{equation}
where $\lVert .\rVert_{c}$ is the vector-wise norm of a matrix across each column vector.

Second, the low-rank matrices $V_{up}$ and $V_{down}$ compress the original weight dimension from $k$ to $r$, which may cause important information in the features to be lost.
Therefore, Graph-LoRA uses PiSSA \cite{meng2024pissa} to initialize the two low-rank matrices $V_{up}$ and $V_{down}$ in order for preserving the most important $r$ features.
Specifically, PiSSA performs singular value decomposition on the original $V$ and takes the first $r$ principal singular components. The first $r$ components represent the $r$ most important features. The low-rank matrices $V_{up}$ and $V_{down}$ are initialized to the same subspace based on the singular value decomposition results, so that their product is initially along the dominant direction of the LLM features. The specific operation is as follows:
\begin{equation}
V = USX^T
\end{equation}
\begin{equation}
V_{down} = U_{[:,\,:r]}\, S^{1/2}_{[:r,\,:r]}\quad\quad
V_{up} = S^{1/2}_{[:r,\,:r]}\, X^{\top}_{[:,\,:r]}
\end{equation}
where $U \in R^{d \times min(d,k)}$ and $X \in R^{k \times min(d,k)}$ are the singular vectors with orthonormal columns, $S \in R^{min(d,k) \times min(d,k)}$ is a diagonal matrix with the singular values arranged in descending order on the diagonal, and $X^T$ is the transpose of $X$. The matrix slicing notations are the same as those in PyTorch, and $[:r]$ and $[:]$ denote the first \emph{r} dimensions and all dimensions respectively.

Third, based on the APSG features generated by GNN, Graph-LoRA merges the graph information in APSG with the sequence information
in LLMs through the attention mechanism. Graph-LoRA aims to use the information in APSG to help LLMs assess the correctness of patches. The specific operations are as follows:
\begin{equation}
F_{APSG} = GNN(F_{APSG})
\end{equation}
\begin{equation}
\Delta V = V_{down}Attention(E, F_{APSG})V_{up}
\end{equation}
where $GNN$ represents the graph neural network, $F_{APSG}$ represents the APSG features of patch, and $E$ represents the sequence features of patch. Specifically, the attention mechanism use weight matrix $W_Q$ to map external information (APSG features of patch, $F_{APSG}$) to query vectors and use weight matrices $W_K$ and $W_V$ to map internal information (sequence features of patch, $E$) to key vectors and value vectors respectively. The attention scores are computed by taking the dot product between the query vectors and key vectors, followed by a softmax that normalizes these attention scores into a probability distribution. The computed attention scores highlight the most relevant positions and are then used to form a weighted sum of the value vectors, so that the model focuses on the most relevant external information. Yao et al. \cite{yao2020multimodal} found that attention can guide the original features to acquire external information. Inspired by this work, we use the $W_Q$ in the multi-head attention mechanism to introduce the graph information of APSG and guide the update of patch features in the LLM. The detailed operations are as follows:}
\begin{equation}
\text{Attention}(E, F_{APSG}) = \text{Concat}(\text{head}_1, \ldots, \text{head}_h) W_O
\end{equation}
\begin{equation}
\text{head}_i(E, F_{APSG}) = S(E, F_{APSG})(E W_V ) 
\end{equation}
\begin{equation}
S(E, F_{APSG})= \text{softmax}\left( \frac{(F_{APSG} W_Q) (E W_K^T)}{\sqrt{d_k}} \right)
\end{equation}
where $W_O$ is a weight matrix. During fine-tuning, the parameters of LLMs are updated via $\Delta m$ and $\Delta V$. The specific operations for updating LLM parameters are as follows:}\\
\begin{equation}
W' = (m + \Delta m)\,\frac{V + \Delta V}{\left\lVert V + \Delta V \right\rVert_{c}}
\end{equation}

Finally, after obtaining the output of patch features by the last layer of the LLM, we use the softmax function as a classifier to assess the correctness of the patch. If the probability of the patch being correct in the classifier output is higher than the probability of overfitting, then the patch is correct, otherwise it is overfitting.

\subsection{Training and Inference}
During training, the parameters in Graph-LoRA and GNN are trained jointly. The additional training parameters of Graph-LoRA are equivalent to 0.6\% of the LLM parameters of 7B size, thus keeping the training cost low. Following the previous studies \cite{zhang2024appt, ye2021automated, lin2022context}, we perform 10-fold cross-validation and take the average of 10 rounds of each training and testing process as the final performance of our method. We use the cross-entropy loss to calculate the gap between the model results and the true value, which has been widely used in classification tasks and previous APCA task. We continuously reduce the gap between the true label y and the model prediction result p to update the model parameters. The cross-entropy loss operation is as follows: 
\begin{equation}
\mathcal{L} = - \left( y \log(p) + (1 - y) \log(1 - p) \right)
\end{equation}

During inference, we first use static analysis techniques to analyze the patch and its context code and obtain the APSG representation corresponding to the patch. Second, we use the trained GNN and LLM to encode the APSG and patch code respectively, obtaining the patch graph features and patch sequence features. Third, the trained Graph-LoRA merges the graph features with the sequence features. Finally, the output of patch features by the LLM is sent to the classifier to assess the correctness of the patch.

\section{Experimental Setup}
To demonstrate the effectiveness of our approach, we design experiments to evaluate the performance of our model. In this section, we introduce the experimental setup. 

\subsection{Research Questions}
For reasonably analyzing the model performance, we explore the following research questions:

\vspace{0.5mm}
\noindent\textbf{RQ1 (Effectiveness): How does our model perform compared to other existing works on the APCA task?} To pursue this question, we evaluate the model on five APCA datasets and compare the performance with that of the state-of-the-art APCA methods.

\vspace{0.5mm}
\noindent\textbf{RQ2 (Impact analysis): How much influence does each part of the model have on the final result?} We gradually remove submodules from the model to evaluate the contribution of each submodule.

\vspace{0.5mm}
\noindent\textbf{RQ3 (Cross-project effectiveness): How does the model perform on patches it has not seen?} We evaluate the model performance in a cross-project prediction scenario to measure the ability of the model to assess new patches. \\
\subsection{Datasets}
In this study, we focus on APCA task datasets constructed with patches
for real-world projects. 
\begin{table}[]
\centering
\caption{\label{tab:t2}Datasets used in our experiment.}
\resizebox{0.48\textwidth}{!}{
\begin{tabular}{llccc}
\hline
Datasets                                                    & Benchmarks                                                               & \multicolumn{1}{l}{\# Correct} & \multicolumn{1}{l}{\# Overfitting} & \multicolumn{1}{l}{Total} \\ \hline
Wang                                                        & Defects4J V1.2                                                           & 248                            & 654                                & 902                       \\ \hline
Merge                                                       & Defects4J V2.0                                                           & 271                            & 2,489                              & 2,760                     \\ \hline
Balance                                                     & Defects4J V2.0                                                           & 271                            & 271                                & 542                       \\ \hline
Lin                                                         & Defects4J V2.0                                                           & 535                            & 648                                & 1,183                     \\ \hline
\begin{tabular}[c]{@{}l@{}}Multi-\\ Benchmarks\end{tabular} & \begin{tabular}[c]{@{}l@{}}Defects4J V2.0,\\ Bus.jar, Bears\end{tabular} & 2673                           & 5121                              & 7794                      \\ \hline
\end{tabular}
}
\end{table}
More specifically, we select five APCA task datasets and Table \ref{tab:t2} gives a summary of these datasets. These datasets range in size from small to large, vary from one to multiple in terms of the number of bug benchmarks used for constructing the datasets, and vary from balanced to imbalanced in terms of the ratio between the number of correct patches and the number of overfitting patches. Next, we give a brief description of these five datasets.

\noindent\textbf{Wang dataset.} The Wang dataset is the most widely used dataset for the APCA task. Wang et al. \cite{wang2020automated} use 21 mainstream APR tools to fix bugs in Defects4J V1.2 and collect the patches generated by these tools. They then check the collected patches and manually assess their correctness. For the patches that pass the tests, they mark them either as correct or overfitting. Finally, they obtain a total of 902 patches, including 248 correct patches and 654 overfitting patches. 

\noindent\textbf{Merge dataset.} The Merge dataset is the largest manually labeled dataset for the APCA task. Yang et al. \cite{yang2023large} manually label the 1,988 patches generated by the PraPR repair system \cite{10.1145/3293882.3330559} and merge them with the Wang dataset by carefully removing the duplicates. They finally obtain 2,760 patches, including 2,489 overfitting patches and 271 correct patches.

\noindent\textbf{Balance dataset.} The Balance dataset contains an equal number of overfitting patches and correct patches. For more thorough evaluations, Yang et al. \cite{yang2023large} construct the Balance dataset to address the problem that the number of correct patches is too different from that of the overfitting patches in the Merge dataset. More specifically, they keep all correct patches from the Merge dataset and sample the same number of overfitting patches from the Merge dataset. 

\noindent\textbf{Lin dataset.} Compared with the Wang dataset, the Lin dataset contains more patches. To better explore the overfitting problem, Lin et al. \cite{lin2022context} add 1,000 patches released by Tian et al. \cite{tian2020evaluating} to the Wang dataset. These 1,000 patches include patches generated by well-known APR tools such as JAID, SketchFix, CapGen, SOFix, and SequenceR, as well as patches written by Defects4J developers. To avoid data leakage, they remove duplicate patches, ultimately obtaining 1183 patches.

\noindent\textbf{Multi-Benchmarks dataset.} 
Since the above four datasets are all constructed with patches for Defects4J, to enable more comprehensive evaluation, we construct a new large patch dataset that involves patches for three bug benchmarks: Defects4J \cite{just2014defects4j}, Bugs.jar \cite{saha2018bugs}, and Bears \cite{madeiral2019bears}. The specific construction process is as follows. First, we add data from the Lin and Merge datasets (note that these two datasets contain all patches from the Wang and Balance datasets) to the Multi-Benchmarks dataset. Then, to avoid an excessive number of overfitting patches for Defects4J, we add human-written patches for Defects4J as the correct patches to the Multi-Benchmarks dataset. Next, to supplement the patches for bug benchmarks besides Defects4J, we augment the Multi-Benchmarks dataset with the patches for bug benchmarks Bugs.jar and Bears released by Ye et al. \cite{ye2021automated}. The patches released by Ye et al. \cite{ye2021automated} consist of human-written patches and patches generated by 11 APR tools. Finally, after deduplication, the Multi-Benchmarks dataset contains 7794 patches, including 2673 correct patches and 5121 overfitting patches. Overall, the patches in the Multi-Benchmarks dataset were both human-written and automatically generated by 22 different APR tools. Note that Lin et al. \cite{lin2022context} and Ye et al. \cite{ye2021automated} similarly use both human-written and automatically generated patches to evaluate APCA methods. Table \ref{tab:t13} summarizes the Multi-Benchmarks dataset.

Overall, the five selected datasets include the largest manually
labeled dataset---the Merge dataset, the relatively small but most widely used dataset---the Wang dataset, the imbalanced datasets---the Wang dataset and the Merge dataset, the (relatively) balanced datasets---the Lin dataset and the Balance dataset, and the large dataset that involves multiple bug benchmarks---the Multi-Benchmarks dataset. Collectively, using these five datasets enables us to conduct a thorough and comprehensive evaluation.
\begin{table}[]
\caption{\label{tab:t13}Summary of the Multi-Benchmarks dataset.}
\resizebox{0.48\textwidth}{!}{
\begin{tabular}{llccc}
\hline
Benchmark                  & Subjects                                            & Correct & Overfitting & All  \\ \hline
\multirow{3}{*}{Defects4J} & Merge dataset \cite{yang2023large} & 271     & 2489        & 2760 \\
                           & Lin dataset \cite{lin2022context}  & 535     & 648         & 1183 \\
                           & Human-written                                       & 798     & 0           & 798  \\ \hline
Bugs.jar                   & Ye et al. \cite{ye2021automated}   & 986     & 2275        & 3261 \\ \hline
Bears                      & Ye et al. \cite{ye2021automated}   & 219     & 531         & 750  \\ \hline
\multicolumn{2}{l}{Total}                                                        & 2809    & 5943        & 8752 \\ \hline
\multicolumn{2}{l}{Total (deduplicated)}                                         & 2673    & 5121        & 7794 \\ \hline
\end{tabular}
}
\end{table}

\subsection{Baselines}\label{baseline}
Following existing studies \cite{yang2023large,zhang2024appt,lin2022context}, our study selects existing state-of-the-art techniques that are designed for or can be adapted to the APCA task. The selected techniques can be broadly categorized into two categories, including \emph{static} and \emph{dynamic} techniques. In particular, we obey the flowing two strategies widely used by existing works \cite{yang2023large,zhang2024appt,lin2022context}. First, we only include techniques that do not rely on the ground-truth patches (\emph{i.e.}, the oracle information) since the ground-truth patch information is unavailable for real-world bug fixing \cite{wang2020automated,overfittingsurvey,yang2023large}.
Second, we only include techniques designed for Java language as Java is the most targeted language in the APR community and the existing patches of real-world bugs are usually available in Java language.

Among the dynamic methods, we select two representative works as our baselines: PATCH-SIM \cite{xiong2018identifying} and Opad \cite{yang2017better}. PATCH-SIM exploits the behavior
similarity of test case executions, and is currently the best among dynamic methods. Opad uses fuzz testing to generate new test cases for exposing overfitting patches, and Opad is further divided into E-Opad and R-Opad according to the different test generation tools used (Evosuite vs Randoop). 



Static methods can be further divided into three categories: traditional methods (denoted as \emph{static-traditional}), machine learning based (ML-based) methods (denoted as \emph{static-ML}), and large language model based (LLM-based) methods (denoted as \emph{static-LLM}). For traditional methods, we consider S3 \cite{le2017s3}, ssFix \cite{weimer2013leveraging}, and CapGen \cite{wen2018context}. For ML-based methods, we consider ODS \cite{ye2021automated}, BERT-LR \cite{tian2020evaluating}, BATS \cite{tian2022predicting}, PANTHER \cite{tian2023best}, CACHE \cite{lin2022context}, and APPT \cite{zhang2024appt}. Moreover, we consider an LLM-based method LLM4PatchCorrect \cite{zhou2024leveraging}. Among traditional methods, S3 assesses the correctness of patches based on six features, ssFix assesses the patch correctness based on structural and conceptual information, and finally CapGen assesses patches based on contextual information. Among the ML-based methods, ODS extracts 202 patch features through abstract syntax trees to describe the correct patch, CACHAE considers both the changed code segments and the related unchanged code segments, and APPT adopts a pre-trained model as an encoder stack and then uses an LSTM stack and a deep learning classifier to evaluate patch correctness. The other three ML-based methods are proposed by Tian and his co-authors. In particular, BERT-LR uses representation learning techniques to learn code change embeddings to assess patch correctness, BATS is an unsupervised method that predicts patch correctness by checking the behaviors of patches against the specifications of failing tests, and PANTHER investigates the advantages of learning code representation and evaluates the correctness of patches by integrating learned embeddings with engineered features. LLM4PatchCorrect is the state-of-the-art method based on LLMs, which predicts patch correctness by feeding an LLM with information of labeled patches, such as error descriptions and failed tests. 

Regarding the results of baselines, 
we reuse the results of baselines from recently published works \cite{zhang2024appt,yang2023large,ye2021automated,zhou2024leveraging} to facilitate a fair comparison whenever it is possible. For experimental completeness, we also reproduce several state-of-the-art APCA methods on the studied datasets that were not considered in previous works. Specifically, we refer to the results by Yang et al. \cite{yang2023large} about dynamic methods PATCH-SIM, E-Opad, and R-Opad and traditional methods S3, ssFix, and CapGen (of the static method category) for relevant datasets.
For ML-based method ODS, the work by Yang et al. \cite{yang2023large} 
reports results for the Wang dataset and the work by Zhang et al. \cite{zhang2024appt} reports results for the Lin dataset. For ML-based methods CACHE and APPT, the work by Zhang et al. \cite{zhang2024appt} reports results for the Lin dataset. For LLM-based method LLM4PatchCorrect, the work by Zhou et al. \cite{zhou2024leveraging} reports results for the Lin dataset using the LLM StarCoder. Likewise, we refer to these results for the purpose of fair comparison.

We next give how we choose APCA methods to reproduce and obtain their results on the studied datasets that were not considered in previous works. For dynamic methods, they have limited practical applications due to the disadvantage of requiring significant time to generate and/or execute tests \cite{zhang2024appt,zhou2024leveraging}. Besides, Patch-SIM can only be applied to Defects4J v1.2 (note that besides the Wang dataset, the other 4 studied datasets all involve Defects4J v2.0) and both Patch-SIM and Opad have a low recall for overfitting patches \cite{yang2023large, wang2020automated}. Thus, we do not reproduce these two dynamic methods. For static methods, previous works have shown that ML-based methods significantly outperform traditional methods \cite{zhang2024appt} and LLM-based methods outperform traditional methods and ML-based methods \cite{zhou2024leveraging}. 
In accordance with this, we consider BERT-RL, BATS, PANTHER, Cache, APPT, and LLM4PatchCorrect to be the most advanced APCA methods proposed recently. To ensure the comprehensiveness of our experimental results, we reproduce these advanced APCA methods on the studied datasets that were not considered in previous works. Specifically, since BERT-RL, BATS, and PANTHER have not been considered by previous works for the studied datasets, we strictly reproduce them according to the corresponding artifacts and obtain their performance on the five datasets. We also reproduce Cache, APPT, and StarCoder-based LLM4PatchCorrect and obtain results on the four studied datasets (excluding the Lin dataset) not considered by previous works.

To give a more extensive evaluation, we account for two other representative LLMs besides StarCoder (used by LLM4PatchCorrect in \cite{zhou2024leveraging}) when a method involves LLMs: CodeLlama and Llama3. More specifically, we implement two variants of LLM4PatchCorrect (\emph{i.e.}, CodeLlama-based LLM4PatchCorrect and Llama3-based LLM4PatchCorrect) and evaluate them on the five studied datasets. We also implement our method using the three LLMs StarCoder, CodeLlama, Llama3, and conduct the evaluation on the five studied datasets.

\subsection{Metrics}
To comprehensively assess the experimental results, we account for multiple evaluation metrics, including accuracy, precision, recall, and \emph{F1} score. Given \emph{TP} that denotes the number of overfitting patches correctly identified as overfitting, \emph{FP} that denotes the number of truly correct patches identified as overfitting, \emph{FN} that refers to the number of overfitting patches identified as correct, and \emph{TN} that refers to the number of truly correct patches identified as correct, these metrics are calculated as follows: 
\begin{itemize} 
\item Accuracy: the ratio of the number of correct predictions to the number of all predictions, given by $(TP+TN)/(TP+FP+TN+FN)$.
\item Precision: the ratio of actual overfitting patches to the overfitting patches predicted by the model, given by $TP/(TP+FP)$.
\item Recall: the ratio of the number of predicted overfitting patches to the number of actual overfitting patches, given by $TP/(TP+FN)$.
\item F1-score: the metric that weighs the accuracy and recall, given by $2*(Precision*Recall)/(Precision+Recall)$.
\end{itemize}

\subsection{Implementation Details}
Our model is implemented using the Pytorch \cite{pytorch} framework. Following the previous APCA work, we use the Adam optimizer \cite{kingma2014adam} to update the model parameters. As the training process progresses, the learning rate is adjusted (ranging from 0 to 0.00005) to adapt to the model learning at different stages. The maximum sequence length is set to be 1024, and words outside the range are ignored. Besides, the dimension of low-rank matrices in Graph-LoRA are set to be 256. Our implementation and evaluation are performed on an Ubuntu 22.04.5 server equipped with two RTX A6000 GPUs.

\section{Experimental Result}
In this section, we present the experimental results for the three research questions.

\subsection{(RQ1) Model Effectiveness}
We compare our method with the selected baselines in the APCA field using the Wang, Merge, Balance, Lin, and Multi-Benchmarks datasets, and the results are shown in Table \ref{tab:t3}, Table \ref{tab:t5}, Table \ref{tab:t4}, Table \ref{tab:t6}, and Table \ref{tab:t14} respectively. For the results of the baselines, we get them according to the way described in Section~\ref{baseline} and the ‘-’ symbol in the tables indicates that the result has not been reported in previous works and we also do not reproduce the corresponding APCA method for reasons given in Section~\ref{baseline}. Since both LLM4PatchCorrect and our approach obtain the best result when Llama3 LLM is used, we refer to the results obtained using Llama3 LLM below when mentioning the results of LLM4PatchCorrect and our approach. However, note that the results obtained using the other two LLMs (CodeLlama and StarCoder) show similar trend. 

Table \ref{tab:t3} shows the results obtained for the Wang dataset. From the results, we can see that our method outperforms all static methods using the four metrics on the Wang dataset. In particular, compared with the APPT method (the state-of-the-art ML-based method), our method is 6.4\%, 6.9\%, 2.2\%, and 4.7\% higher in accuracy, precision, recall, and F1 score respectively. Compared with the LLM4PatchCorrect method (the most advanced LLM-based method), our method is 3.4\%, 3.5\%, 2.5\%, and 3.0\% higher in accuracy, precision, recall, and F1 score respectively. Among the dynamic methods, Opad relies on a large number of generated test cases to achieve better results in precision and it takes a lot of time to assess patches. Currently, our method is the closest to Opad among static methods, and it significantly outperforms all dynamic methods in comprehensive evaluation metrics such as F1 score. This result suggests that our method better captures important information for patch correctness prediction and improves the performance of LLMs on the APCA task. 
\begin{table}[]
\centering
\caption{\label{tab:t3} Effectiveness comparison on the Wang dataset.}
\resizebox{0.5\textwidth}{!}{
\begin{tabular}{llcccc}
\hline
Category                                                                        & Method                                                                & Accuracy                      & Precision                     & Recall                        & F1                            \\ \hline
                                                                                & PATCH-SIM                                                             & 49.5\%                        & 83.0\%                        & 38.9\%                        & 53.0\%                        \\ \cline{2-6} 
                                                                                & E-Opad                                                                & 34.9\%                        & \textbf{100.0\%}              & 10.2\%                        & 18.5\%                        \\ \cline{2-6} 
\multirow{-3}{*}{Dynamic}                                                       & R-Opad                                                                & 37.7\%                        & \textbf{100.0\%}              & 14.7\%                        & 25.6\%                        \\ \hline
                                                                                & S3                                                                    & 69.6\%                        & 79.1\%                        & 79.1\%                        & 79.1\%                        \\ \cline{2-6} 
                                                                                & ssFix                                                                 & 69.2\%                        & 78.8\%                        & 78.8\%                        & 78.8\%                        \\ \cline{2-6} 
\multirow{-3}{*}{\begin{tabular}[c]{@{}l@{}}Static-\\ traditional\end{tabular}} & CapGen                                                                & 68.1\%                        & 78.0\%                        & 78.0\%                        & 78.0\%                        \\ \hline
                                                                                & ODS                                                                   & 88.9\%                        & 90.4\%                        & 94.8\%                        & 92.5\%                        \\ \cline{2-6} 
                                                                                &  BERT\_LR                                       &  89.4\% &  90.7\% &  95.1\% &  92.5\% \\ \cline{2-6} 
                                                                                &  BATS                                          &  89.8\% &  90.8\% &  95.6\% &  93.1\%\\ \cline{2-6} 
                                                                                &  PANTHER                                        &  91.0\% &  91.3\% &  95.3\% &  93.3\% \\ \cline{2-6} 
                                                                                & Cache                                                                 & 90.1\%                        & 91.2\%                        & 94.5\%                        & 92.9\%                        \\ \cline{2-6} 
\multirow{-6}{*}{Static-ML}                                                     & APPT                                                                  & 90.4\%                        & 91.5\%                        & 96.0\%                        & 93.6\%                        \\ \hline
                                                                                & \begin{tabular}[c]{@{}l@{}}LLM4PatchCorrect-\\ CodeLlama\end{tabular} & 92.4\%                        & 93.7\%                        & 94.7\%                        & 94.2\%                        \\ \cline{2-6} 
                                                                                & \begin{tabular}[c]{@{}l@{}}LLM4PatchCorrect-\\ StarCoder\end{tabular} & 92.6\%                        & 93.9\%                        & 94.8\%                        & 94.4\%                        \\ \cline{2-6} 
\multirow{-3}{*}{Static-LLM}                                                    & \begin{tabular}[c]{@{}l@{}}LLM4PatchCorrect-\\ Llama3\end{tabular}    & 93.4\%                        & 94.9\%                        & 95.7\%                        & 95.3\%                        \\ \hline
                                                                                & \begin{tabular}[c]{@{}l@{}}Graph-LoRA-\\ CodeLlama\end{tabular}       & \textbf{95.8\%}               & 96.9\%                        & \textbf{97.6\%}               & \textbf{97.3\%}               \\ \cline{2-6} 
                                                                                & \begin{tabular}[c]{@{}l@{}}Graph-LoRA-\\ StarCoder\end{tabular}       & \textbf{96.1\%}               & 96.9\%                        & \textbf{97.8\%}               & \textbf{97.4\%}               \\ \cline{2-6} 
\multirow{-3}{*}{Our}                                                           & \begin{tabular}[c]{@{}l@{}}Graph-LoRA-\\ Llama3\end{tabular}          & \textbf{96.8\%}               & 98.4\%                        & \textbf{98.2\%}               & \textbf{98.3\%}               \\ \hline
\end{tabular}
}
\end{table}


\begin{table}[]
\centering
\caption{\label{tab:t5}Effectiveness comparison on the Merge dataset.}
\resizebox{0.5\textwidth}{!}{
\begin{tabular}{llcccc}
\hline
Category                                                                        & Method                                                                & Accuracy                      & Precision                     & Recall                        & F1                            \\ \hline
                                                                                & PATCH-SIM                                                             & -                             & \textbf{-}                    & -                             & -                             \\ \cline{2-6} 
                                                                                & E-Opad                                                                & 22.2\%                        & \textbf{99.4\%}               & 13.8\%                        & 24.2\%                        \\ \cline{2-6} 
\multirow{-3}{*}{Dynamic}                                                       & R-Opad                                                                & 24.1\%                        & 96.5\%                        & 16.4\%                        & 28.0\%                        \\ \hline
                                                                                & S3                                                                    & 82.6\%                        & 90.4\%                        & 90.4\%                        & 90.4\%                        \\ \cline{2-6} 
                                                                                & ssFix                                                                 & 82.1\%                        & 90.1\%                        & 90.1\%                        & 90.1\%                        \\ \cline{2-6} 
\multirow{-3}{*}{\begin{tabular}[c]{@{}l@{}}Static-\\ traditional\end{tabular}} & CapGen                                                                & 82.9\%                        & 90.5\%                        & 90.5\%                        & 90.5\%                        \\ \hline
                                                                                & ODS                                                                   & -                             & -                             & -                             & -                             \\ \cline{2-6} 
                                                                                &  BERT\_LR                                       &  90.9\% &  91.3\% &  91.0\% &  91.1\% \\ \cline{2-6} 
                                                                                &  BATS                                           &  91.4\% &  91.2\% &  91.4\% &  91.3\% \\ \cline{2-6} 
                                                                                &  PANTHER                                        &  92.1\% &  92.1\% &  91.7\% &  91.9\% \\ \cline{2-6} 
                                                                                & Cache                                                                 & 91.7\%                        & 91.9\%                        & 90.1\%                        & 91.8\%                        \\ \cline{2-6} 
\multirow{-6}{*}{Static-ML}                                                     & APPT                                                                  & 92.2\%                        & 92.5\%                        & 92.1\%                        & 92.3\%                        \\ \hline
                                                                                & \begin{tabular}[c]{@{}l@{}}LLM4PatchCorrect-\\ CodeLlama\end{tabular} & 93.2\%                        & 94.1\%                        & 92.9\%                        & 93.5\%                        \\ \cline{2-6} 
                                                                                & \begin{tabular}[c]{@{}l@{}}LLM4PatchCorrect-\\ StarCoder\end{tabular} & 93.6\%                        & 94.3\%                        & 93.1\%                        & 93.7\%                        \\ \cline{2-6} 
\multirow{-3}{*}{Static-LLM}                                                    & \begin{tabular}[c]{@{}l@{}}LLM4PatchCorrect-\\ Llama3\end{tabular}    & 94.5\%                        & 95.1\%                        & 94.2\%                        & 94.6\%                        \\ \hline
                                                                                & \begin{tabular}[c]{@{}l@{}}Graph-LoRA-\\ CodeLlama\end{tabular}       & \textbf{96.5\%}               & 96.8\%                        & \textbf{96.6\%}               & \textbf{96.7\%}               \\ \cline{2-6} 
                                                                                & \begin{tabular}[c]{@{}l@{}}Graph-LoRA-\\ StarCoder\end{tabular}       & \textbf{96.7\%}               & 96.8\%                        & \textbf{96.6\%}               & \textbf{96.7\%}               \\ \cline{2-6} 
\multirow{-3}{*}{Our}                                                           & \begin{tabular}[c]{@{}l@{}}Graph-LoRA-\\ Llama3\end{tabular}          & \textbf{97.6\%}               & 97.6\%                        & \textbf{97.9\%}               & \textbf{97.8\%}               \\ \hline
\end{tabular}
}
\end{table}

Table \ref{tab:t5} shows the results obtained for the Merge dataset. From table \ref{tab:t5}, we can see that our method outperforms all baselines in terms of accuracy, recall and F1 score and outperforms all baselines except E-Opad in terms of precision. Compared with the APPT method, our method is 5.4\%, 5.1\%, 5.8\%, and 5.5\% higher in accuracy, precision, recall, and F1 score respectively. Compared with the LLM4PatchCorrect method, our method is 3.1\%, 2.5\%, 3.7\%, and 3.2\% higher in accuracy, precision, recall, and F1 score respectively. This result again proves that our method can achieve excellent performance in the accurate manually labeled dataset.
\begin{table}[]
\centering
\caption{\label{tab:t4} Effectiveness comparison on the Balance dataset.}
\resizebox{0.5\textwidth}{!}{
\begin{tabular}{llcccc}
\hline
Category                                                                        & Method                                                                & Accuracy                      & Precision                     & Recall                        & F1                            \\ \hline
                                                                                & PATCH-SIM                                                             & -                             & \textbf{-}                    & -                             & -                             \\ \cline{2-6} 
                                                                                & E-Opad                                                                & 58.5\%                        & \textbf{96.0\%}               & 17.7\%                        & 29.9\%                        \\ \cline{2-6} 
\multirow{-3}{*}{Dynamic}                                                       & R-Opad                                                                & 55.4\%                        & 74.6\%                        & 16.2\%                        & 26.7\%                        \\ \hline
                                                                                & S3                                                                    & 44.3\%                        & 44.3\%                        & 44.3\%                        & 44.3\%                        \\ \cline{2-6} 
                                                                                & ssFix                                                                 & 46.5\%                        & 46.5\%                        & 46.5\%                        & 46.5\%                        \\ \cline{2-6} 
\multirow{-3}{*}{\begin{tabular}[c]{@{}l@{}}Static-\\ traditional\end{tabular}} & CapGen                                                                & 49.1\%                        & 49.1\%                        & 49.1\%                        & 49.1\%                        \\ \hline
                                                                                & ODS                                                                   & -                             & -                             & -                             & -                             \\ \cline{2-6} 
                                                                                &  BERT\_LR                                       &  63.4\% &  61.7\% &  63.4\% &  62.5\% \\ \cline{2-6} 
                                                                                &  BATS                                           &  64.2\% &  63.2\% &  64.7\% &  63.9\% \\ \cline{2-6} 
                                                                                &  PANTHER                                        &  69.5\% &  67.6\% &  69.7\% &  68.6\% \\ \cline{2-6} 
                                                                                & Cache                                                                 & 68.6\%                        & 69.5\%                        & 67.3\%                        & 68.4\%                        \\ \cline{2-6} 
\multirow{-6}{*}{Static-ML}                                                     & APPT                                                                  & 71.8\%                        & 72.7\%                        & 73.6\%                        & 73.1\%                        \\ \hline
                                                                                & \begin{tabular}[c]{@{}l@{}}LLM4PatchCorrect-\\ CodeLlama\end{tabular} & 75.4\%                        & 75.8\%                        & 76.4\%                        & 75.9\%                        \\ \cline{2-6} 
                                                                                & \begin{tabular}[c]{@{}l@{}}LLM4PatchCorrect-\\ StarCoder\end{tabular} & 75.7\%                        & 76.0\%                        & 76.7\%                        & 76.3\%                        \\ \cline{2-6} 
\multirow{-3}{*}{Static-LLM}                                                    & \begin{tabular}[c]{@{}l@{}}LLM4PatchCorrect-\\ Llama3\end{tabular}    & 79.2\%                        & 80.1\%                        & 80.8\%                        & 80.4\%                        \\ \hline
                                                                                & \begin{tabular}[c]{@{}l@{}}Graph-LoRA-\\ CodeLlama\end{tabular}       & \textbf{82.5\%}               & 83.8\%                        & \textbf{82.6\%}               & \textbf{83.2\%}               \\ \cline{2-6} 
                                                                                & \begin{tabular}[c]{@{}l@{}}Graph-LoRA-\\ StarCoder\end{tabular}       & \textbf{82.8\%}               & 84.2\%                        & \textbf{83.1\%}               & \textbf{83.6\%}               \\ \cline{2-6} 
\multirow{-3}{*}{Our}                                                           & \begin{tabular}[c]{@{}l@{}}Graph-LoRA-\\ Llama3\end{tabular}          & \textbf{86.7\%}               & 87.8\%                        & \textbf{87.2\%}               & \textbf{87.5\%}               \\ \hline
\end{tabular}
    }
\end{table}

\begin{table}[]
\centering
\caption{\label{tab:t6}Effectiveness comparison on the Lin dataset.}
\resizebox{0.5\textwidth}{!}{
\begin{tabular}{llcccc}
\hline
Category                     & Method                                                                & Accuracy                      & Precision                     & Recall                        & F1                            \\ \hline
                             & ODS                                                                   & 62.3\%                        & 68.5\%                        & 69.7\%                        & 69.1\%                        \\ \cline{2-6} 
                             &  BERT\_LR                                       &  68.5\% &  71.6\% &  73.4\% &  72.5\% \\ \cline{2-6} 
                             &  BATS                                           &  68.9\% &  72.8\% &  74.6\% &  73.7\% \\ \cline{2-6} 
                             &  PANTHER                                        &  72.6\% &  74.1\% &  78.6\% &  76.4\% \\ \cline{2-6} 
                             & CACHE                                                                 & 75.4\%                        & 79.5\%                        & 76.5\%                        & 78.0\%                        \\ \cline{2-6} 
\multirow{-6}{*}{Static-ML}  & APPT                                                                  & 79.7\%                        & 80.8\%                        & 83.2\%                        & 81.8\%                        \\ \hline
                             & \begin{tabular}[c]{@{}l@{}}LLM4PatchCorrect-\\ CodeLlama\end{tabular} & 83.7\%                        & 86.8\%                        & 87.7\%                        & 87.2\%                        \\ \cline{2-6} 
                             & \begin{tabular}[c]{@{}l@{}}LLM4PatchCorrect-\\ StarCoder\end{tabular} & 84.0\%                        & 87.1\%                        & 87.9\%                        & 87.5\%                        \\ \cline{2-6} 
\multirow{-3}{*}{Static-LLM} & \begin{tabular}[c]{@{}l@{}}LLM4PatchCorrect-\\ Llama3\end{tabular}    & 86.2\%                        & 88.4\%                        & 89.0\%                        & 88.7\%                        \\ \hline
                             & \begin{tabular}[c]{@{}l@{}}Graph-LoRA-\\ CodeLlama\end{tabular}       & \textbf{90.7\%}               & \textbf{90.7\%}               & \textbf{90.2\%}               & \textbf{90.5\%}               \\ \cline{2-6} 
                             & \begin{tabular}[c]{@{}l@{}}Graph-LoRA-\\ StarCoder\end{tabular}       & \textbf{90.7\%}               & \textbf{90.4\%}               & \textbf{89.8\%}               & \textbf{90.1\%}               \\ \cline{2-6} 
\multirow{-3}{*}{Our}        & \begin{tabular}[c]{@{}l@{}}Graph-LoRA-\\ Llama3\end{tabular}          & \textbf{91.8\%}               & \textbf{92.5\%}               & \textbf{92.2\%}               & \textbf{92.3\%}               \\ \hline
\end{tabular}
}
\end{table}

Table \ref{tab:t4} shows the results obtained for the Balance dataset. With regard to this dataset, our method still outperforms all baseline methods in terms of accuracy, recall, and F1 score, and is also better than all baseline methods except E-Opad in terms of precision. Notably, our method has a more obvious improvement on the Balance dataset than the Wang dataset. Compared with the APPT method, our method improves accuracy, precision, recall, and F1 score by 15.7\%, 15.1\%, 13.6\%, and 14.4\% respectively. Compared with the LLM4PatchCorrect method, our method improves accuracy, precision, recall, and F1 score by 7.5\%, 7.7\%, 6.4\%, and 7.1\% respectively. This demonstrates that our method is more suitable for the situation where the number of positive samples and that of negative samples are balanced. 
\begin{table}[]
\caption{\label{tab:t14}Effectiveness comparison on the Multi-Benchmarks dataset.}
\resizebox{0.5\textwidth}{!}{

\begin{tabular}{llcccc}
\hline
\textbf{Category}           & \textbf{Method}                                                       & \textbf{Accuracy} & \textbf{Precision} & \textbf{Recall} & \textbf{F1}     \\ \hline
\multirow{5}{*}{Static-ML}  & BERT\_LR                                                              & 80.6\%            & 79.6\%             & 83.3\%          & 81.4\%          \\ \cline{2-6} 
                            & BATS                                                                  & 82.1\%            & 82.4\%             & 84.2\%          & 83.3\%          \\ \cline{2-6} 
                            & PANTHER                                                               & 85.0\%            & 84.2\%             & 85.0\%          & 84.6\%          \\ \cline{2-6} 
                            & CACHE                                                                 & 82.8\%            & 82.0\%             & 84.1\%          & 83.0\%          \\ \cline{2-6} 
                            & APPT                                                                  & 85.6\%            & 84.8\%             & 85.5\%          & 85.1\%          \\ \hline
\multirow{3}{*}{Static-LLM} & \begin{tabular}[c]{@{}l@{}}LLM4PatchCorrect-\\ CodeLlama\end{tabular} & 89.0\%            & 88.2\%             & 89.5\%          & 88.8\%          \\ \cline{2-6} 
                            & \begin{tabular}[c]{@{}l@{}}LLM4PatchCorrect-\\ StarCoder\end{tabular} & 89.4\%            & 88.8\%             & 90.0\%          & 89.4\%          \\ \cline{2-6} 
                            & \begin{tabular}[c]{@{}l@{}}LLM4PatchCorrect-\\ Llama3\end{tabular}    & 90.2\%            & 89.4\%             & 90.6\%          & 90.0\%          \\ \hline
\multirow{3}{*}{Our}        & \begin{tabular}[c]{@{}l@{}}Graph-LoRA-\\ CodeLlama\end{tabular}       & \textbf{93.0\%}   & \textbf{92.4\%}    & \textbf{93.4\%} & \textbf{92.9\%} \\ \cline{2-6} 
                            & \begin{tabular}[c]{@{}l@{}}Graph-LoRA-\\ StarCoder\end{tabular}       & \textbf{93.6\%}   & \textbf{92.7\%}    & \textbf{93.6\%} & \textbf{93.1\%} \\ \cline{2-6} 
                            & \begin{tabular}[c]{@{}l@{}}Graph-LoRA-\\ Llama3\end{tabular}          & \textbf{94.6\%}   & \textbf{93.3\%}    & \textbf{94.8\%} & \textbf{94.0\%} \\ \hline
\end{tabular}
}
\end{table}

Table \ref{tab:t6} shows the results obtained for the Lin dataset. Note that the work by Yang et al. \cite{yang2023large} does not use this dataset, so we do not have results for some selected baselines. 
Similarly, we can see from the table that our method outperforms all APCA baseline methods based on machine learning and LLMs. Compared with the APPT method, our method outperforms it by 12.6\%, 11.7\%, 9.0\%, and 10.5\% in accuracy, precision, recall, and F1 score respectively. Compared with the LLM4PatchCorrect method, our method outperforms it by 6.1\%, 4.1\%, 3.2\%, and 3.6\% in accuracy, precision, recall, and F1 score respectively. Note that compared with the Wang dataset and the Merge dataset, this dataset is more balanced and our method again has a more obvious improvement.

Table \ref{tab:t14} shows the results obtained for the newly constructed Multi-Benchmarks dataset. It can be seen from the table that our method outperforms all baselines,  achieving 94.6\%, 93.3\%, 94.5\%, and 93.0\% in terms of accuracy, precision, recall, and F1 score respectively. Compared with the APPT method, our method outperforms it by 9.0\%, 8.8\%, 9.0\%, and 8.8\% in accuracy, precision, recall, and F1 score respectively. Compared with the LLM4PatchCorrect method, our method outperforms it by 4.4\%, 3.9\%, 4.2\%, and 4.0\% in accuracy, precision, recall, and F1 score respectively. Overall, the result again demonstrates that our method can achieve that best performance on the APCA task. 
\begin{tcolorbox}[width=\linewidth,boxrule=0pt,top=1pt, bottom=1pt, left=1pt,right=1pt, colback=gray!20,colframe=gray!20]
\textbf{Answer to RQ1:} Overall, our experimental results show that: (1) Our method outperforms all static APCA methods in all metrics and datasets; (2) Compared with the state-of-the-art APCA method LM4PatchCorrect, our method improves the accuracy, precision, recall and F1 score by 3.1\% to 7.5\%, 2.5\% to 7.7\%, 2.6\% to 6.4\%, and 3.0\% to 7.1\% respectively; (3) Our method achieves better performance for the situation where the number of correct patches and that of the overfitting patches are balanced. 
\end{tcolorbox}

\subsection{(RQ2) Ablation Study}\label{ablationstudy}
To demonstrate the effectiveness of each sub-element and show its contribution to the final results, we perform two  ablation studies using the three considered LLMs Llama3, StarCoder, and CodeLlama. In addition, considering the ratio between the number of correct patches and the number of overfitting patches, the ablation studies are conducted using the Wang dataset (\emph{imbalanced}) and the Balance dataset (\emph{balanced}).

The first ablation study focuses on the model structure to demonstrate the contributions of the sub-modules in the model to the final results, and the second ablation study focuses on APSG nodes to demonstrate the contributions of different nodes in APSG to the final results. For the first ablation study, we start with the complete model and then gradually remove specific components and observe the results after removal. Specifically, to observe the role of the attention mechanism, we first remove the attention fusion layer of Graph-LoRA and replace it with Graph-LoRA-Weak which achieves fusion through vector concatenation. To observe whether GNNs are more effective in acquiring graph information than linearizing the graph, we then remove Graph-LoRA-Weak and directly input the linearized APSG content into the LLM in the form of sequences. Next, we delete the attributes of APSG and only input the graph structure of APSG and code patch into the LLM to observe the role of the patch attributes. After that, we remove the whole APSG and only input the code patch into the LLM to observe the role of the graph structure information of APSG. Finally, we do not train the model and only give the LLMs the prompt "You are a model responsible for assessing patch correctness. Assess whether the patch is correct." 
to prove the effectiveness of training. For the second ablation study, we gradually remove the variable nodes, control nodes, and context nodes from the complete model and observe the experimental results. Since APSG is built around patches, we do not remove the patch nodes to preserve the meaning of APSG.
The obtained results for the first ablation study are shown in Tables \ref{tab:t7} and  \ref{tab:t8}, and the obtained results for the second ablation study are shown in Tables \ref{tab:t15} and \ref{tab:t18}. 

\begin{table}[]
\centering
\caption{\label{tab:t7}Ablation Study for model structure on the Wang dataset.}
\resizebox{0.5\textwidth}{!}{
\begin{tabular}{lcccc}
\hline
Model                                   & Accuracy                      & Precision                     & Recall                        & F1                            \\ \hline
Graph-LoRA-Llama3                       & \textbf{97.3\%}               & \textbf{98.6\%}               & \textbf{98.4\%}               & \textbf{98.5\%}               \\
-Graph-LoRA-Attention                   & 96.6\%                        & 97.6\%                        & 97.2\%                        & 97.4\%                        \\
-Graph-LoRA-Weak                        & 95.1\%                        & 95.1\%                        & 95.4\%                        & 95.3\%                        \\
-APSG-Attribute                         & 94.1\%                        & 94.3\%                        & 94.5\%                        & 94.4\%                        \\
-APSG-Graph                             & 92.7\%                        & 92.2\%                        & 92.0\%                        & 92.1\%                        \\
 -Llama3-Train    &  30.8\% &  14.3\% &  35.6\% &  20.4\% \\ \hline
Graph-LoRA-StarCoder                    & \textbf{96.5\%}               & \textbf{97.5\%}               & \textbf{98.2\%}               & \textbf{97.8\%}               \\
-Graph-LoRA-Attention                   & 95.2\%                        & 96.4\%                        & 97.0\%                        & 96.6\%                        \\
-Graph-LoRA-Weak                        & 93.6\%                        & 94.3\%                        & 94.5\%                        & 94.4\%                        \\
-APSG-Attribute                         & 92.7\%                        & 93.2\%                        & 93.6\%                        & 93.4\%                        \\
-APSG-Graph                             & 90.8\%                        & 91.0\%                        & 91.2\%                        & 91.1\%                        \\
 -StarCoder-Train &  28.3\% &  10.7\% &  2.4\%  &  3.9\%  \\ \hline
Graph-LoRA-CodeLlama                    & \textbf{96.2\%}               & \textbf{97.2\%}               & \textbf{97.8\%}               & \textbf{97.5\%}               \\
-Graph-LoRA-Attention                   & 95.3\%                        & 96.2\%                        & 96.8\%                        & 96.5\%                        \\
-Graph-LoRA-Weak                        & 93.6\%                        & 94.2\%                        & 94.4\%                        & 94.3\%                        \\
-APSG-Attribute                         & 92.7\%                        & 93.0\%                        & 93.3\%                        & 93.2\%                        \\
-APSG-Graph                             & 90.6\%                        & 90.2\%                        & 90.5\%                        & 90.4\%                        \\
 -CodeLlama-Train &  17.6\% &  1.3\%  &  1.7\%  &  1.5\%  \\ \hline
\end{tabular}
}
\end{table}

\begin{table}[]
\centering
\caption{\label{tab:t8}Ablation Study for model structure on the Balance dataset.}
\resizebox{0.5\textwidth}{!}{
\begin{tabular}{lcccc}
\hline
\textbf{Model}                          & \textbf{Accuracy}             & \textbf{Precision}           & \textbf{Recall}               & \textbf{F1}                  \\ \hline
Graph-LoRA-Llama3                       & \textbf{86.7\%}               & \textbf{87.8\%}              & \textbf{87.2\%}               & \textbf{87.5\%}              \\
-Graph-LoRA-Attention                   & 85.5\%                        & 86.8\%                       & 86.1\%                        & 86.4\%                       \\
-Graph-LoRA-Weak                        & 83.2\%                        & 84.1\%                       & 83.6\%                        & 83.8\%                       \\
-APSG-Attribute                         & 82.3\%                        & 83.5\%                       & 82.8\%                        & 83.1\%                       \\
-APSG-Graph                             & 79.7\%                        & 81.0\%                       & 80.6\%                        & 80.8\%                       \\
 -Llama3-Train    &  21.6\% &  4.1\% &  20.7\% &  6.8\% \\ \hline
Graph-LoRA-StarCoder                    & \textbf{82.8\%}               & \textbf{84.2\%}              & \textbf{83.1\%}               & \textbf{83.6\%}              \\
-Graph-LoRA-Attention                   & 81.4\%                        & 82.6\%                       & 81.8\%                        & 82.1\%                       \\
-Graph-LoRA-Weak                        & 78.7\%                        & 79.6\%                       & 78.9\%                        & 79.2\%                       \\
-APSG-Attribute                         & 78.0\%                        & 78.8\%                       & 78.3\%                        & 78.4\%                       \\
-APSG-Graph                             & 75.3\%                        & 75.8\%                       & 75.3\%                        & 75.5\%                       \\
 -StarCoder-Train &  18.3\% & 2.3\% &  1.5\%  &  1.8\% \\ \hline
Graph-LoRA-CodeLlama                    & \textbf{82.5\%}               & \textbf{83.8\%}              & \textbf{82.6\%}               & \textbf{83.2\%}              \\
-Graph-LoRA-Attention                   & 81.0\%                        & 82.2\%                       & 81.1\%                        & 81.6\%                       \\
-Graph-LoRA-Weak                        & 78.3\%                        & 79.2\%                       & 78.5\%                        & 78.7\%                       \\
-APSG-Attribute                         & 77.2\%                        & 78.1\%                       & 77.5\%                        & 77.6\%                       \\
-APSG-Graph                             & 74.9\%                        & 75.7\%                       & 75.0\%                        & 75.2\%                       \\
 -CodeLlama-Train &  14.2\% &  1.6\% &  1.3\%  &  1.5\% \\ \hline
\end{tabular}
}
\end{table}
For the first ablation study, we can have the following observations. First, the model performance  decreases after removing the attention mechanism within Graph-LoRA. In the imbalanced Wang dataset, the model performance decreases by 0.7\% to 1.1\%, 0.9\% to 1.1\%, 1.0\% to 1.2\%, and 1.0\% to 1.2\% in terms of the accuracy, precision, recall, and F1 score respectively. In the balanced Balance dataset, the model performance in terms of the accuracy, precision, recall, and F1 score decreases by 1.2\% to 1.5\%, 1.0\% to 1.6\%, 1.1\% to 1.5\%, and 1.1\% to 1.6\% respectively. This shows that compared to directly concatenating graph features and text features, the attention mechanism can better help LLMs acquire graph information. 

Second, after removing Graph-LoRA-Weak, the model performance  decreases significantly. In the imbalanced Wang dataset, the model performance decreases by 1.5\% to 1.7\%, 1.8\% to 2.5\%, 1.8\% to 2.5\%, and 2.1\% to 2.2\% in terms of the accuracy, precision, recall, and F1 score respectively. In the balanced Balance dataset, the model performance in terms of the accuracy, precision, recall, and F1 score decreases by 2.3\% to 2.7\%, 2.7\% to 3.0\%, 2.5\% to 2.9\%, and 2.6\% to 2.9\% respectively. This shows that compared to inputting linearized graph information into LLM, GNNs can obtain graph information more effectively. 
\begin{table}[]
\caption{\label{tab:t15}Ablation Study for APSG nodes on the Wang
dataset.}
\resizebox{0.5\textwidth}{!}{
\begin{tabular}{lcccc}
\hline
\textbf{Model}       & \textbf{Accuracy} & \textbf{Precision} & \textbf{Recall} & \textbf{F1}     \\ \hline
Graph-LoRA-Llama3    & \textbf{97.3\%}   & \textbf{98.6\%}    & \textbf{98.4\%} & \textbf{98.5\%} \\
-Variable Node       & 96.2\%            & 97.3\%             & 97.4\%          & 97.3\%          \\
-Control Node        & 95.9\%            & 97.1\%             & 97.0\%          & 97.1\%          \\
-Context Node        & 93.9\%            & 95.2\%             & 94.6\%          & 94.9\%          \\ \hline
Graph-LoRA-StarCoder & \textbf{96.5\%}   & \textbf{97.5\%}    & \textbf{98.2\%} & \textbf{97.8\%} \\
-Variable Node       & 95.3\%            & 96.4\%             & 97.0\%          & 96.7\%          \\
-Control Node        & 95.0\%            & 95.9\%             & 96.6\%          & 96.2\%          \\
-Context Node        & 92.6\%            & 93.2\%             & 93.5\%          & 93.3\%          \\ \hline
Graph-LoRA-CodeLlama & \textbf{96.2\%}   & \textbf{97.2\%}    & \textbf{97.8\%} & \textbf{97.5\%} \\
-Variable Node       & 94.8\%            & 96.0\%             & 96.7\%          & 96.3\%          \\
-Control Node        & 94.3\%            & 95.6\%             & 96.1\%          & 95.8\%          \\
-Context Node        & 92.2\%            & 92.8\%             & 93.0\%          & 92.9\%          \\ \hline
\end{tabular}
}
\end{table}

\begin{table}[]
\caption{\label{tab:t18}Ablation Study for APSG nodes on the Balance dataset.}
\resizebox{0.5\textwidth}{!}{
\begin{tabular}{lcccc}
\hline
\textbf{Model}       & \textbf{Accuracy} & \textbf{Precision} & \textbf{Recall} & \textbf{F1}     \\ \hline
Graph-LoRA-Llama3    & \textbf{86.7\%}   & \textbf{87.8\%}    & \textbf{87.2\%} & \textbf{87.5\%} \\
-Variable Node       & 85.3\%            & 86.3\%             & 85.9\%          & 86.1\%          \\
-Control Node        & 85.1\%            & 86.0\%             & 85.7\%          & 85.8\%          \\
-Context Node        & 83.0\%            & 83.6\%             & 82.7\%          & 83.1\%          \\ \hline
Graph-LoRA-StarCoder & \textbf{82.8\%}   & \textbf{84.2\%}    & \textbf{83.1\%} & \textbf{83.6\%} \\
-Variable Node       & 81.1\%            & 82.4\%             & 81.3\%          & 81.8\%          \\
-Control Node        & 80.5\%            & 81.9\%             & 80.8\%          & 81.3\%          \\
-Context Node        & 77.2\%            & 78.5\%             & 77.8\%          & 78.1\%          \\ \hline
Graph-LoRA-CodeLlama & \textbf{82.5\%}   & \textbf{83.8\%}    & \textbf{82.6\%} & \textbf{83.2\%} \\
-Variable Node       & 80.8\%            & 82.0\%             & 80.9\%          & 81.4\%          \\
-Control Node        & 80.3\%            & 81.6\%             & 80.3\%          & 80.9\%          \\
-Context Node        & 76.8\%            & 78.1\%             & 77.1\%          & 77.6\%          \\ \hline
\end{tabular}
}
\end{table}
Third, after deleting the attribute of APSG and keeping only the graph structure of APSG, the model performance  also drops. In the imbalanced Wang dataset, the model performance drops by 0.9\% to 1.0\%, 0.8\% to 1.4\%, 0.9\% to 1.1\%, and 0.9\% to 1.1\% in terms of the accuracy, precision, recall, and F1 scores respectively. In the balanced Balance dataset, the model performance in terms of the accuracy, precision, recall, and F1 score decreases by 0.7\% to 1.1\%, 0.6\% to 1.1\%, 0.6\% to 1.0\%, and 0.7\% to 1.1\% respectively. This suggests that explicit code attributes can help LLM determine the correctness of patches. 

Fourth, the performance of the model again obviously decreases if the whole APSG is deleted. In the imbalanced Wang dataset, the performance of the model decreases by 1.4\% to 2.1\%, 2.1\% to 2.8\%, 2.4\% to 2.8\% and 2.7\% to 2.8\% in terms of the accuracy, precision, recall, and F1 score, respectively. In the balanced Balance dataset, the performance of the model in terms of the accuracy, precision, recall, and F1 score decreases by 2.3\% to 2.7\%, 2.4\% to 2.9\%, 2.2\% to 3.0\%, and 2.3\% to 2.9\% respectively. This shows that the graph structure information of APSG, which captures the patch semantic through data and control flow between program elements, is vital for helping LLM assess the correctness of the patches.

Finally, if the model is given only the LLM prompt "You are a model responsible for assessing patch correctness. Assess whether the patch is correct.", the performance of all three LLMs drops significantly on both datasets, especially for the LLMs StarCoder and CodeLlama. This result proves the effectiveness of our training process. At the same time, the low-quality results of the basic LLMs also suggest that the three LLMs we used likely have no data leakage issue on the studied datasets, and our method is the primary reason for the improvement in model performance.

For the second ablation study, the following observations can be made. First, the model performance obviously decreases after removing the variable nodes. In the imbalanced Wang dataset, the model performance decreases by 1.1\% to 1.4\%, 1.1\% to 1.3\%, 1.0\% to 1.2\% and 1.1\% to 1.2\% in terms of the accuracy, precision, recall, and F1 score respectively. In the balanced Balance dataset, the model performance with respect to the accuracy, precision, recall, and F1 score decreases by 1.4\% to 1.7\%, 1.5\% to
1.8\%, 1.3\% to 1.8\%, and 1.4\% to 1.8\% respectively. This shows that variable nodes provide important information in assessing patch correctness. Second, the model performance decreases slightly after removing the control nodes. In the imbalanced Wang dataset, the model performance decreases by 0.3\% to 0.5\%, 0.2\% to 0.5\%, 0.4\% to 0.6\% and 0.2\% to 0.5\% in terms of the accuracy, precision, recall, and F1 score respectively. In the balanced Balance dataset, the model performance in terms of the accuracy, precision, recall, and F1 score decreases by 0.2\% to 0.6\%, 0.3\% to
0.5\%, 0.2\% to 0.6\%, and 0.3\% to 0.5\% respectively. We believe that the slight decrease in model performance is due to the relatively small proportion of control nodes. Nevertheless, the control nodes still obviously enrich the patch semantic information. Finally, the model performance decreases significantly after removing the context nodes in APSG. In the imbalanced Wang dataset, the model performance decreases by 2.0\% to 2.4\%, 1.9\% to 2.8\%, 2.4\% to 3.1\% and 2.2\% to 2.9\% in terms of the accuracy, precision, recall, and F1 score respectively. In the balanced Balance dataset, the model performance with respect to the accuracy, precision, recall, and F1 score decreases by 2.1\% to 3.5\%, 2.4\% to
3.5\%, 3.0\% to 3.2\%, and 2.7\% to 3.3\% respectively. This result demonstrates that the context nodes store vital semantic information and serve as an important basis for assessing patch correctness.
\begin{table}[]
\centering
\caption{\label{tab:t9}Effectiveness of our method in a cross-project setting on the Wang dataset.}
\resizebox{0.5\textwidth}{!}{
\begin{tabular}{clcccc}
\hline
\multicolumn{1}{l}{Project} & Approach         & Accuracy        & Precision       & Recall          & F1              \\ \hline
\multirow{3}{*}{Chart}      & APPT             & 82.2\%          & 82.7\%          & 84.2\%          & 83.9\%          \\ \cline{2-6} 
                            & LLM4PatchCorrect & 90.3\%          & 90.5\%          & 91.3\%          & 90.8\%          \\ \cline{2-6} 
                            & Our              & \textbf{92.3\%} & \textbf{93.6\%} & \textbf{92.8\%} & \textbf{93.3\%} \\ \hline
\multirow{3}{*}{Closure}    & APPT             & 77.2\%          & 78.4\%          & 81.6\%          & 81.5\%          \\ \cline{2-6} 
                            & LLM4PatchCorrect & 85.4\%          & 88.2\%          & 89.5\%          & 88.9\%          \\ \cline{2-6} 
                            & Our              & \textbf{88.7\%} & \textbf{90.6\%} & \textbf{91.4\%} & \textbf{91.0\%} \\ \hline
\multirow{3}{*}{Lang}       & APPT             & 80.7\%          & 79.6\%          & 80.8\%          & 80.2\%          \\ \cline{2-6} 
                            & LLM4PatchCorrect & 89.1\%          & 88.7\%          & 90.3\%          & 89.5\%          \\ \cline{2-6} 
                            & Our              & \textbf{91.9\%} & \textbf{92.1\%} & \textbf{92.6\%} & \textbf{92.3\%} \\ \hline
\multirow{3}{*}{Math}       & APPT             & 80.4\%          & 82.7\%          & 84.6\%          & 83.4\%          \\ \cline{2-6} 
                            & LLM4PatchCorrect & 90.4\%          & 90.6\%          & 91.3\%          & 90.9\%          \\ \cline{2-6} 
                            & Our              & \textbf{93.2\%} & \textbf{93.4\%} & \textbf{93.0\%} & \textbf{93.2\%} \\ \hline
\multirow{3}{*}{Time}       & APPT             & 87.4\%          & 83.9\%          & 80.8\%          & 84.4\%          \\ \cline{2-6} 
                            & LLM4PatchCorrect & 94.8\%          & 93.5\%          & 94.9\%          & 94.2\%          \\ \cline{2-6} 
                            & Our              & \textbf{95.9\%} & \textbf{96.0\%} & \textbf{96.5\%} & \textbf{96.2\%} \\ \hline
\multirow{3}{*}{Average}    & APPT             & 81.6\%          & 81.5\%          & 82.4\%          & 82.7\%          \\ \cline{2-6} 
                            & LLM4PatchCorrect & 90.0\%          & 90.3\%          & 91.1\%          & 90.9\%          \\ \cline{2-6} 
                            & Our              & \textbf{92.4\%} & \textbf{93.1\%} & \textbf{93.3\%} & \textbf{93.2\%} \\ \hline
\end{tabular}
}
\end{table}
\begin{table}[]
\centering
\caption{\label{tab:t10}Effectiveness of our method in a cross-project setting on the Merge dataset.}
\resizebox{0.5\textwidth}{!}{
\begin{tabular}{clcccc}
\hline
\multicolumn{1}{l}{Project} & Approach         & Accuracy        & Precision       & Recall          & F1              \\ \hline
\multirow{3}{*}{Chart}      & APPT             & 86.1\%          & 86.7\%          & 87.2\%          & 86.9\%          \\ \cline{2-6} 
                            & LLM4PatchCorrect & 91.5\%          & 91.8\%          & 90.7\%          & 91.2\%          \\ \cline{2-6} 
                            & Our              & \textbf{93.4\%} & \textbf{93.7\%} & \textbf{93.1\%} & \textbf{93.4\%} \\ \hline
\multirow{3}{*}{Closure}    & APPT             & 81.8\%          & 82.5\%          & 83.1\%          & 82.8\%          \\ \cline{2-6} 
                            & LLM4PatchCorrect & 87.5\%          & 88.2\%          & 88.7\%          & 88.4\%          \\ \cline{2-6} 
                            & Our              & \textbf{89.8\%} & \textbf{91.3\%} & \textbf{90.9\%} & \textbf{91.1\%} \\ \hline
\multirow{3}{*}{Lang}       & APPT             & 83.5\%          & 82.6\%          & 83.7\%          & 83.1\%          \\ \cline{2-6} 
                            & LLM4PatchCorrect & 90.7\%          & 90.5\%          & 91.5\%          & 91.0\%          \\ \cline{2-6} 
                            & Our              & \textbf{92.8\%} & \textbf{93.4\%} & \textbf{93.0\%} & \textbf{93.2\%} \\ \hline
\multirow{3}{*}{Math}       & APPT             & 86.1\%          & 87.7\%          & 88.3\%          & 88.0\%          \\ \cline{2-6} 
                            & LLM4PatchCorrect & 91.6\%          & 91.3\%          & 90.1\%          & 90.7\%          \\ \cline{2-6} 
                            & Our              & \textbf{94.0\%} & \textbf{94.3\%} & \textbf{93.7\%} & \textbf{94.0\%} \\ \hline
\multirow{3}{*}{Time}       & APPT             & 88.7\%          & 87.1\%          & 86.2\%          & 86.6\%          \\ \cline{2-6} 
                            & LLM4PatchCorrect & 94.8\%          & 93.1\%          & 94.8\%          & 93.9\%          \\ \cline{2-6} 
                            & Our              & \textbf{96.9\%} & \textbf{97.4\%} & \textbf{96.0\%} & \textbf{96.8\%} \\ \hline
\multirow{3}{*}{Average}    & APPT             & 85.2\%          & 85.3\%          & 85.7\%          & 85.5\%          \\ \cline{2-6} 
                            & LLM4PatchCorrect & 91.2\%          & 91.0\%          & 91.2\%          & 91.0\%          \\ \cline{2-6} 
                            & Our              & \textbf{93.4\%} & \textbf{94.0\%} & \textbf{93.3\%} & \textbf{93.6\%} \\ \hline
\end{tabular}
}
\end{table}
\begin{tcolorbox}[width=\linewidth,boxrule=0pt,top=1pt, bottom=1pt, left=1pt,right=1pt, colback=gray!20,colframe=gray!20]
\textbf{Answer to RQ2:} The performance of the model after removing different elements in two ablation studies shows that: (1) all the main components of the proposed method contribute positively to the final results, especially the graph part of
APSG and the GNN part of Graph-LoRA; (2) different types of nodes in APSG all contribute to the final result to varying degrees, and the context nodes have the most obvious impact.
\end{tcolorbox}

\begin{table}[]
\centering
\caption{\label{tab:t11}Effectiveness of our method in a cross-project setting on the Balance dataset.}
\resizebox{0.5\textwidth}{!}{
\begin{tabular}{clcccc}
\hline
\multicolumn{1}{l}{Project} & Approach         & Accuracy        & Precision       & Recall          & F1              \\ \hline
\multirow{3}{*}{Chart}      & APPT             & 65.2\%          & 63.8\%          & 66.2\%          & 65.0\%          \\ \cline{2-6} 
                            & LLM4PatchCorrect & 74.6\%          & 76.5\%          & 77.3\%          & 76.9\%          \\ \cline{2-6} 
                            & Our              & \textbf{79.7\%} & \textbf{82.6\%} & \textbf{83.5\%} & \textbf{83.0\%} \\ \hline
\multirow{3}{*}{Closure}    & APPT             & 61.6\%          & 63.7\%          & 67.2\%          & 65.4\%          \\ \cline{2-6} 
                            & LLM4PatchCorrect & 71.3\%          & 73.5\%          & 73.8\%          & 73.6\%          \\ \cline{2-6} 
                            & Our              & \textbf{77.3\%} & \textbf{79.2\%} & \textbf{80.4\%} & \textbf{79.8\%} \\ \hline
\multirow{3}{*}{Lang}       & APPT             & 65.3\%          & 66.4\%          & 67.1\%          & 66.7\%          \\ \cline{2-6} 
                            & LLM4PatchCorrect & 70.8\%          & 72.6\%          & 73.3\%          & 72.9\%          \\ \cline{2-6} 
                            & Our              & \textbf{78.7\%} & \textbf{82.9\%} & \textbf{83.4\%} & \textbf{83.1\%} \\ \hline
\multirow{3}{*}{Math}       & APPT             & 63.6\%          & 66.2\%          & 69.8\%          & 68.0\%          \\ \cline{2-6} 
                            & LLM4PatchCorrect & 74.1\%          & 75.3\%          & 75.8\%          & 75.5\%          \\ \cline{2-6} 
                            & Our              & \textbf{81.3\%} & \textbf{80.4\%} & \textbf{80.9\%} & \textbf{80.7\%} \\ \hline
\multirow{3}{*}{Time}       & APPT             & 70.2\%          & 68.4\%          & 60.4\%          & 64.1\%          \\ \cline{2-6} 
                            & LLM4PatchCorrect & 77.2\%          & 77.8\%          & 78.1\%          & 77.9\%          \\ \cline{2-6} 
                            & Our              & \textbf{80.9\%} & \textbf{82.7\%} & \textbf{83.1\%} & \textbf{82.9\%} \\ \hline
\multirow{3}{*}{Average}    & APPT             & 65.2\%          & 65.7\%          & 66.1\%          & 66.0\%          \\ \cline{2-6} 
                            & LLM4PatchCorrect & 73.6\%          & 75.1\%          & 75.7\%          & 75.4\%          \\ \cline{2-6} 
                            & Our              & \textbf{79.6\%} & \textbf{81.6\%} & \textbf{82.3\%} & \textbf{81.9\%} \\ \hline
\end{tabular}
}
\end{table}

\begin{table}[]
\centering
\caption{\label{tab:t12}Effectiveness of our method in a cross-project setting on the Lin dataset.}
\resizebox{0.5\textwidth}{!}{
\begin{tabular}{clcccc}
\hline
\multicolumn{1}{l}{Project} & Approach         & Accuracy        & Precision       & Recall          & F1              \\ \hline
\multirow{3}{*}{Chart}      & APPT             & 73.5\%          & 71.0\%          & 76.4\%          & 73.6\%          \\ \cline{2-6} 
                            & LLM4PatchCorrect & 87.8\%          & 88.5\%          & 88.7\%          & 88.6\%          \\ \cline{2-6} 
                            & Our              & \textbf{90.5\%} & \textbf{91.2\%} & \textbf{91.4\%} & \textbf{91.3\%} \\ \hline
\multirow{3}{*}{Closure}    & APPT             & 66.9\%          & 69.3\%          & 89.8\%          & 78.2\%          \\ \cline{2-6} 
                            & LLM4PatchCorrect & 80.4\%          & 83.6\%          & 84.2\%          & 83.9\%          \\ \cline{2-6} 
                            & Our              & \textbf{85.4\%} & \textbf{87.6\%} & \textbf{88.3\%} & \textbf{87.9\%} \\ \hline
\multirow{3}{*}{Lang}       & APPT             & 73.0\%          & 83.6\%          & 71.0\%          & 71.3\%          \\ \cline{2-6} 
                            & LLM4PatchCorrect & 83.8\%          & 83.2\%          & 85.3\%          & 84.2\%          \\ \cline{2-6} 
                            & Our              & \textbf{89.1\%} & \textbf{88.7\%} & \textbf{89.6\%} & \textbf{89.1\%} \\ \hline
\multirow{3}{*}{Math}       & APPT             & 69.1\%          & 70.5\%          & 84.3\%          & 76.8\%          \\ \cline{2-6} 
                            & LLM4PatchCorrect & 85.2\%          & 86.9\%          & 87.8\%          & 87.3\%          \\ \cline{2-6} 
                            & Our              & \textbf{89.6\%} & \textbf{89.8\%} & \textbf{90.3\%} & \textbf{90.0\%} \\ \hline
\multirow{3}{*}{Time}       & APPT             & 80.0\%          & 71.4\%          & 66.7\%          & 69.0\%          \\ \cline{2-6} 
                            & LLM4PatchCorrect & 88.5\%          & 89.6\%          & 89.3\%          & 89.4\%          \\ \cline{2-6} 
                            & Our              & \textbf{90.8\%} & \textbf{91.6\%} & \textbf{91.9\%} & \textbf{91.7\%} \\ \hline
\multirow{3}{*}{Average}    & APPT             & 72.5\%          & 70.8\%          & 77.6\%          & 74.1\%          \\ \cline{2-6} 
                            & LLM4PatchCorrect & 85.1\%          & 86.4\%          & 87.1\%          & 86.7\%          \\ \cline{2-6} 
                            & Our              & \textbf{88.5\%} & \textbf{91.6\%} & \textbf{90.3\%} & \textbf{90.9\%} \\ \hline
\end{tabular}
}
\end{table}

\begin{table}[]
\centering
\caption{\label{tab:t17}Effectiveness of our method in a cross-project setting on the Multi-Benchmarks dataset.}
\resizebox{0.5\textwidth}{!}{
\begin{tabular}{clcccc}
\hline
\multicolumn{1}{l}{Project} & Approach         & Accuracy        & Precision       & Recall          & F1              \\ \hline
\multirow{3}{*}{Chart}      & APPT             & 77.5\%          & 75.0\%          & 78.4\%          & 76.7\%          \\ \cline{2-6} 
                            & LLM4PatchCorrect & 89.8\%          & 90.5\%          & 90.7\%          & 90.6\%          \\ \cline{2-6} 
                            & Our              & \textbf{92.3\%} & \textbf{92.4\%} & \textbf{92.8\%} & \textbf{92.6\%} \\ \hline
\multirow{3}{*}{Closure}    & APPT             & 70.9\%          & 73.3\%          & 75.8\%          & 76.9\%          \\ \cline{2-6} 
                            & LLM4PatchCorrect & 82.4\%          & 85.6\%          & 86.2\%          & 85.9\%          \\ \cline{2-6} 
                            & Our              & \textbf{86.9\%} & \textbf{89.1\%} & \textbf{89.8\%} & \textbf{89.4\%} \\ \hline
\multirow{3}{*}{Lang}       & APPT             & 77.3\%          & 77.6\%          & 75.0\%          & 76.3\%          \\ \cline{2-6} 
                            & LLM4PatchCorrect & 85.8\%          & 85.2\%          & 87.3\%          & 86.2\%          \\ \cline{2-6} 
                            & Our              & \textbf{90.6\%} & \textbf{90.2\%} & \textbf{91.1\%} & \textbf{90.6\%} \\ \hline
\multirow{3}{*}{Math}       & APPT             & 73.1\%          & 74.5\%          & 78.3\%          & 76.4\%          \\ \cline{2-6} 
                            & LLM4PatchCorrect & 87.2\%          & 88.9\%          & 89.8\%          & 89.3\%          \\ \cline{2-6} 
                            & Our              & \textbf{91.1\%} & \textbf{91.3\%} & \textbf{91.8\%} & \textbf{91.5\%} \\ \hline
\multirow{3}{*}{Time}       & APPT             & 84.0\%          & 75.4\%          & 70.7\%          & 73.0\%          \\ \cline{2-6} 
                            & LLM4PatchCorrect & 90.5\%          & 91.6\%          & 91.3\%          & 91.4\%          \\ \cline{2-6} 
                            & Our              & \textbf{92.3\%} & \textbf{93.1\%} & \textbf{93.4\%} & \textbf{93.2\%} \\ \hline
\multirow{3}{*}{Average}    & APPT             & 76.6\%          & 75.2\%          & 75.6\%          & 75.9\%          \\ \cline{2-6} 
                            & LLM4PatchCorrect & 87.1\%          & 88.4\%          & 89.1\%          & 88.7\%          \\ \cline{2-6} 
                            & Our              & \textbf{90.6\%} & \textbf{91.2\%} & \textbf{91.8\%} & \textbf{91.5\%} \\ \hline
\end{tabular}
}
\end{table}
\subsection{(RQ3) Cross-Project Prediction}
Through the above experiments, we have demonstrated that the performance of our method for the APCA task is optimal in a cross-validation setting. However, in practical applications, the model needs to assess patches from unseen projects. To further explore the performance of our method, we conduct the cross-project verification using the Wang, Merge, Balance, Lin, and Multi-Benchmarks datasets. For example, with regard to a certain dataset, we use patches from projects other than Chart to train the model and use patches from Chart to evaluate the model. As the three datasets \emph{Merge, Balance, and Multi-Benchmarks} contain patch data from projects other than the five listed projects Chart, Closure, Lang, Math, and Time, we regard these patch data from other projects as training data when conducting cross-project verification on these three datasets. In the experiments, we use two state-of-the-art APCA methods as baseline methods, including APPT and LLM4PatchCorrect. To effectively evaluate the performance upper limit of the APCA methods, our method and the LM4PatchCorrect method are both based on the LLM Llama3. 

Table \ref{tab:t9}, Table \ref{tab:t10}, Table \ref{tab:t11}, Table \ref{tab:t12}, and Table \ref{tab:t17} show the results of cross-project prediction on the Wang, Merge, Balance, Lin, and Multi-Benchmarks datasets respectively. From the tables, we can see that in the cross-project prediction scenario, the accuracy, precision, recall, and F1 score of our method are 92.4\%, 93.1\%, 93.3\%, 93.2\% respectively on the Wang dataset, 93.4\%, 94.0\%, 93.4\%, 93.7\% respectively on the Merge dataset, 79.6\%, 81.6\%, 82.3\%, 81.9\% respectively on the Balance dataset, 88.5\% 91.6\% 90.3\% 90.9\% respectively on the Lin dataset, and 90.6\%, 91.2\%, 91.8\%, 91.5\% respectively on the Multi-Benchmarks dataset. From the results, we can see that the performance of our method has declined in processing unseen patches. However, note that our method still outperforms all APCA baseline models. Compared with the best APPT method (the state-of-the-art ML-based method), our method has improved accuracy, precision, recall, and F1 score by 8.2\% to 16.0\%, 8.7\% to 20.8\%, 7.7\% to 16.2\%, and 8.2\% to 16.8\% respectively. Compared with the LLM4PatchCorrect methdo (the most advanced LLM-based method), our method has improved accuracy, precision, recall, and F1 score by 2.2\% to 6.0\%, 2.8\% to 6.5\%, 2.1\% to 6.6\%, and 2.3\% to 6.5\% respectively.

\begin{tcolorbox}[width=\linewidth,boxrule=0pt,top=1pt, bottom=1pt, left=1pt,right=1pt, colback=gray!20,colframe=gray!20]
\textbf{Answer to RQ3:} The performance under a cross-project
scenario demonstrates that: (1) Compared with the cross-validation setting, the performance of APCA methods in the cross-project scenario generally deteriorates; (2) In the cross-project scenario, our method still achieves the state-of-the-art performance using all metrics and datasets.
\end{tcolorbox}
\begin{figure*}[h]
    \centering
    \subfigure[true negative case]{
        \includegraphics[width=0.9\textwidth]{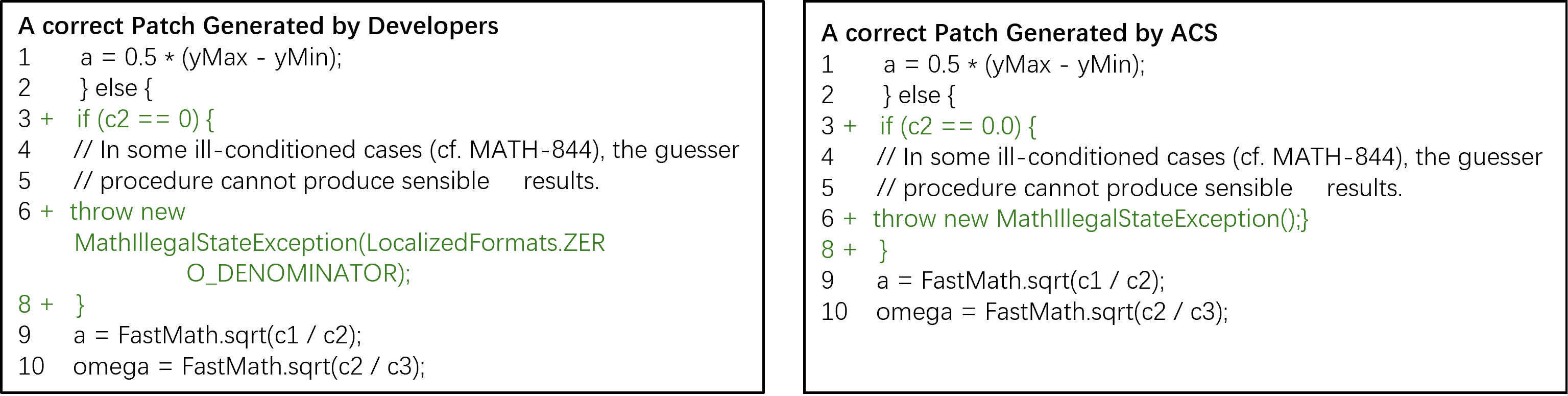}
    }
    \subfigure[false negative case]{
        \includegraphics[width=0.9\textwidth]{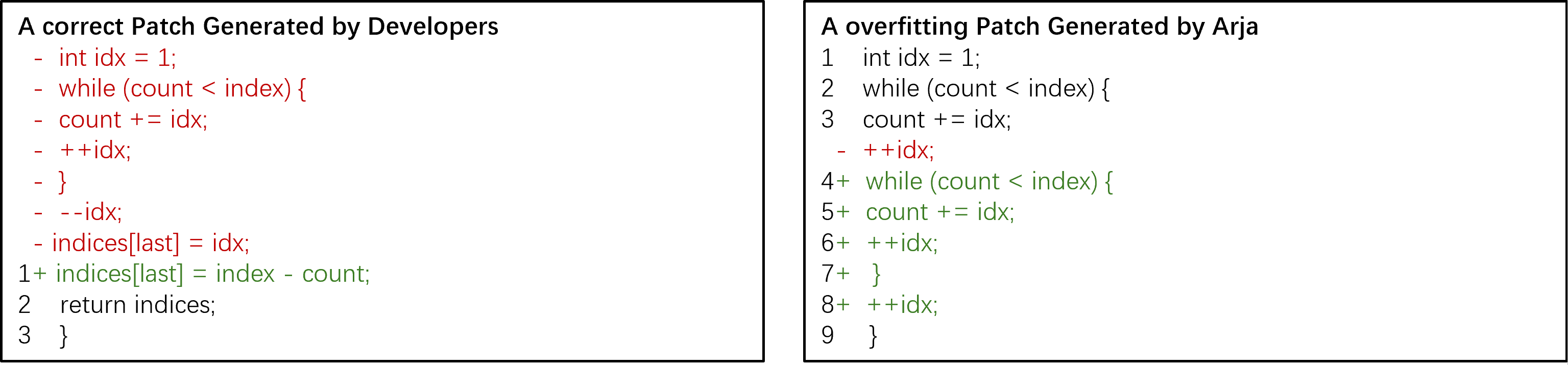}
    }
     \subfigure[false positive case]{
        \includegraphics[width=0.9\textwidth]{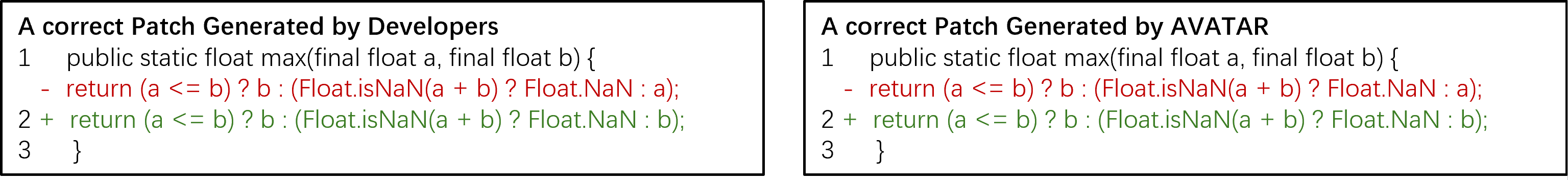}
    }
     \subfigure[true positive case]{
        \includegraphics[width=0.9\textwidth]{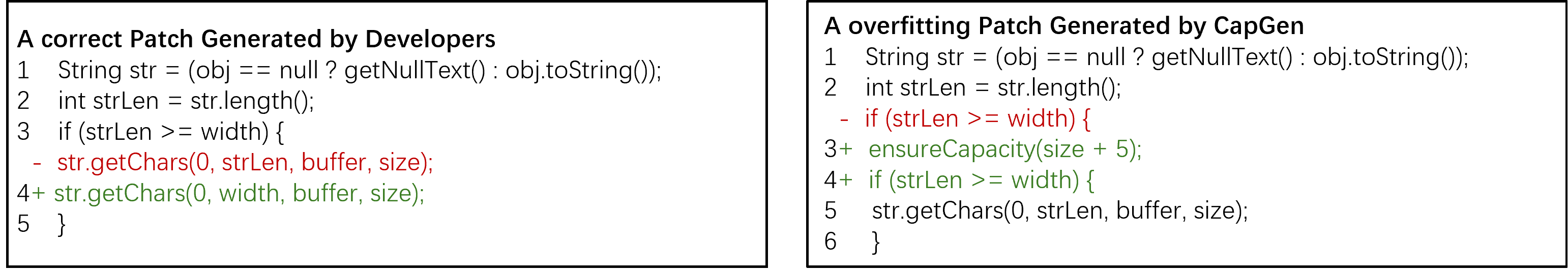}
    }
    \caption{\label{fig:frog4}An overview of the case study.}
    \label{fig:main}
\end{figure*}
\section{Discussion}
\subsection{Case Study}
To reasonably explain how the model works, we conduct a case analysis of the experimental results. We select four specific cases from the experimental results for detailed analysis, including predicting the correct patch as correct, predicting the correct patch as incorrect, predicting the overfitting patch as correct, and predicting the overfitting patch as incorrect. The selected cases are shown in Fig.~\ref{fig:frog4}. 

\vspace{0.5mm}
\textbf{True negative case:} Figure 4(a) shows an example of a correct patch generated for Math-25 by ACS. We find that this patch 1) changes the control flow with the newly introduced branch If, and 2) a frequently used code segment \emph{throw new MathIllegalStateException} appears in line 6, which makes the patch has a high entropy value. Our model captures these features and then considers them similar to the features of the correct patch, thus predicting the generated patch as correct.

\vspace{0.5mm}
\textbf{False negative case:} Figure 4(b) shows an example of an overfitting patch generated for Math-56 by Arja. Both our model and the LLM4PatchCorrect model predict it as a correct patch. We analyze the patch and find that neither of them significantly changes the code semantics. However, we find that line 4 and 5 of the patch generated by Arja are the same as the context code. Since we assume that the context code is correct during training, it affects the judgment of the model.

\vspace{0.5mm}
\textbf{False positive case:} Figure 4(c) shows an example of a correct patch generated for Math-59 by AVATAR. However, our model mistakenly classifies it as an overfitting patch. We note that the defective code lies in a conditional branch, which incorrectly calculates b as a. This patch has less context and is only related with the variable name. The model may not effectively obtain the feature of the variable name, which shows that deep learning models rely on richer context information.

\vspace{0.5mm}
\textbf{True positive case:} Figure 4(d) shows an example of an overfitting patch generated for Lang-59 by CapGen. Our model successfully predicts this patch as an overfitting patch but the LLM4PatchCorrect model does not. We analyze the patch and find that it does not significantly change the code, resulting in less effective information about the patch. Due to the addition of APSG graph features, our model can obtain more patch information and thus more accurately assess the correctness of the patch.

\subsection{Effect of the Ground-Truth Patches}
In this section, we aim to explore the effect of ground-truth patches on the model performance. 
In the APCA task, ground-truth patches are typically used for APR experimentation \cite{yang2023large,overfittingsurvey}, where ground-truth patches may be accessible and plausible patches that behave differently from the ground-truth patches are deemed as overfitting.
However, note that the ground-truth patch information is unavailable for real-world bug fixing, and our method focuses on this scenario and 
does not rely on ground-truth data.
In fact, there are already some APCA works that use ground-truth patches to assess patch correctness. For example, 
Xin et al. \cite{xin2017identifying} propose DiffTGen, which generates new tests based on the execution process of the ground-truth patches to improve the robustness of the method. Ye et al. \cite{ye2021scale} compare the differences in runtime information between ground-truth patches and APR-generated patches to assess patch correctness.

\begin{figure}
\centering
\includegraphics[width=0.48\textwidth]{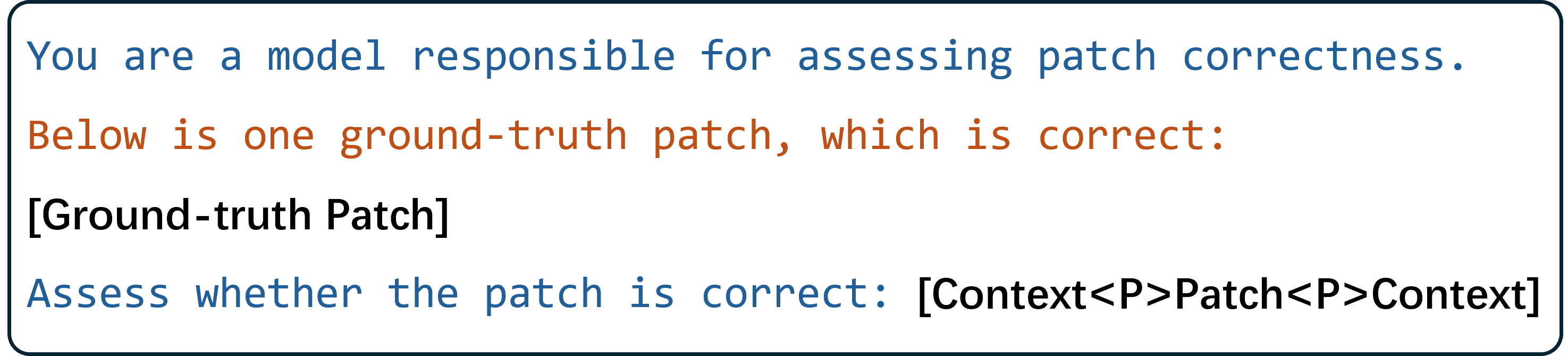}
\caption{\label{fig:frog5}The input with prompt of ground-truth patch.}
\end{figure}

To investigate the effect of ground-truth patches on the model performance, we use them as prompt inputs to the model and conduct experiments on the Lin dataset and the Balance dataset. More specifically, we first collect Defects4J human-written patches as ground-truth patches. Then, for each patch to be assessed, we construct a patch pair consisting of the patch itself and the ground-truth patch for the corresponding bug. For example, in the Lin dataset, there are 17 patches to be assessed for Chart-1. We thus construct 17 patch pairs using each patch to be assessed and the ground-truth 
patch for the corresponding bug. Next, during data preprocessing (\emph{i.e.}, the lower half of step (a) in Fig.~\ref{fig:frog2}), we place the ground-truth patch in the patch pair (preceding the patch to be assessed) as a prompt to feed into the model. The input with prompt of ground-truth patch is shown in Fig.~\ref{fig:frog5}. Finally, we train and test the model with ground-truth prompt in the same experimental settings as RQ1. Note that when evaluating the performance of the model with prompt of ground-truth patch, we remove the human-written patches from the evaluation data to prevent data leakage. The results are given in Table \ref{tab:t16}.


From the table, we can see that in the Balance dataset, the performance of the model with prompt of ground-truth patch improves by 1.5\%, 1.3\%, 1.1\%, and 1.2\% in terms of the accuracy, precision, recall, and F1 score respectively. In the Lin dataset, the performance of the model with prompt of ground-truth patch improves by 0.7\%, 0.8\%, 0.6\%, and 0.7\% in terms of the accuracy, precision, recall, and F1 score respectively. The results suggest that after adding prompt of ground-truth patch, the performance of the model improves by varying degrees on the two datasets. The improvement is more remarkable on the Balance dataset, which contains fewer patch samples. Overall, we believe that ground-truth patches can help the model assess patch correctness, especially when the data is scarce.
\begin{table}[]
\centering
\caption{\label{tab:t16}The result of adding the prompt of ground-truth patch to the model.}
\resizebox{0.5\textwidth}{!}{
\begin{tabular}{llcccc}
\hline
\textbf{Dataset}                                                            & \textbf{Model}                                                             & \textbf{Accuracy} & \textbf{Precision} & \textbf{Recall} & \textbf{F1}     \\ \hline
\multirow{2}{*}{\begin{tabular}[c]{@{}l@{}}Balance \\ dataset\end{tabular}} & Graph-LoRA-Llama3                                                          & 86.7\%            & 87.8\%             & 87.2\%          & 87.5\%          \\ \cline{2-6} 
                                                                            & \begin{tabular}[c]{@{}l@{}}Graph-LoRA-Llama3\\ +Ground-Truth\end{tabular}  & \textbf{88.2\%}   & \textbf{89.1\%}    & \textbf{88.3\%} & \textbf{88.7\%} \\ \hline
\multirow{2}{*}{\begin{tabular}[c]{@{}l@{}}Lin\\ dataset\end{tabular}}      & Graph-LoRA-Llama3                                                          & 91.8\%            & 92.5\%             & 92.2\%          & 92.3\%          \\ \cline{2-6} 
                                                                            & \begin{tabular}[c]{@{}l@{}}Graph-LoRA-Llama3 \\ +Ground-Truth\end{tabular} & \textbf{92.5\%}   & \textbf{93.3\%}    & \textbf{92.8\%} & \textbf{93.0\%} \\ \hline
\end{tabular}
}
\end{table}

\subsection{Threats to Validity}
\textbf{Threats to external validity.} A threat to external validity is related with whether our results can be generalized. To minimize this threat, 1) we conduct experiments on five APCA datasets, ranging in size from large to small, vary from one to multiple in terms of the number of bug benchmarks used for constructing the datasets, and vary from balanced to imbalanced in terms of the ratio between the number of correct patches and the number of overfitting patches; 2) we also consider three different representative LLMs when LLM is involved with in this study; 3) with regard to the baselines, we always select the most representative and state-of-the-art techniques in each category of existing works on APCA task. Another threat to external validity is related with the implementation. Our implementation currently supports Java language only, and further efforts are needed to apply our approach to other programming languages. We consider addressing this limitation as an important direction for future work.

\textbf{Threats to internal validity.} One threat to internal validity is that we can possibly introduce errors during the experimental
process. To reduce this threat as much as possible, several
authors have carefully and independently examined the
artifacts. Besides, to facilitate the replication and verification of our work, we have made the relevant materials (including code, datasets, models, etc.) publicly available for the community to review. Another threat to internal validity concerns the use of LLM, and the issue is that the LLM during the pre-training process may possibly have encountered the content of the used datasets. However, this is a common potential issue faced by most studies that use LLMs for code related tasks. In particular, note that this potential issue is also faced by the baseline method LLM4PatchCorrect in our experiment. Using the same LLM, our method consistently demonstrates clear advantages over LLM4PatchCorrect. This suggests that our method itself offers new insights for statically predicting patch correctness. Meanwhile, the results of the ablation study (specifically the part of discarding training and only giving simple prompt) in Section~\ref{ablationstudy} also suggest that the three LLMs we used likely have no data leakage issue on the data datasets we used. 
\section{Conclusion}
Patch overfitting is a serious issue which overshadows the automated program repair area, and many research efforts have been devoted for automated patch correctness assessment (APCA). With the emergence of large language model (LLM) technology, researchers have employed LLM to assess the patch correctness. The literature on APCA has highlighted the importance of capturing patch semantic and explicit code attributes in predicting patch correctness. However, existing LLM-based methods 1) typically treat code as token sequences and ignore the inherent formal structure for code, and 2) do not explicitly
account for enough code attributes. To overcome these drawbacks, we in this paper design a novel patch graph representation named attributed patch semantic graph (APSG), which adequately captures the patch semantic and explicitly reflects important patch attributes. To effectively use graph information in APSG, we accordingly propose a new parameter-efficient fine-tuning (PEFT) method of LLMs named Graph-LoRA. The results of extensive evaluations show that compared to the state-of-the-art methods, our method improves the accuracy and F1 score by 3.1\% to 7.5\% and 3.0\% to 7.1\% respectively. For future work, we will apply our method to more LLMs and demonstrate the effectiveness of our method in more code related tasks. 
\section*{Acknowledgments}
\noindent
We deeply appreciate the anonymous reviewers for their insightful comments. This work was supported by National Natural Science Foundation of China (Grant No. 62102233), Shandong Province Overseas Outstanding Youth Fund (Grant No. 2022HWYQ-043), Joint Key Funds of National Natural Science Foundation of China (Grant No. U24A20244), and Qilu Young Scholar Program of Shandong University. 

\bibliographystyle{unsrt} 
\bibliography{references} 

\end{document}